\documentclass[aps,prb,twocolumn,groupedaddress,amsmath]{revtex4-2}
\usepackage{graphicx}
\usepackage{bm}
\usepackage{amssymb}
\usepackage[normalem]{ulem}
\usepackage{float}
\usepackage{dcolumn}
\usepackage{subfigure}
\usepackage{threeparttable}
\usepackage{multirow}
\usepackage[usenames,dvipsnames]{color}
\usepackage{mathtools}

\begin{document}
\newcommand{\vn}[1]{{\boldsymbol{#1}}}
\newcommand{\vht}[1]{{\boldsymbol{#1}}}
\newcommand{\matn}[1]{{\bf{#1}}}
\newcommand{\matnht}[1]{{\boldsymbol{#1}}}
\newcommand{\bege}{\begin{equation}}
\newcommand{\ee}{\end{equation}}
\newcommand{\bal}{\begin{aligned}}
\newcommand{\defbar}{\overline}
\newcommand{\SM}{\scriptstyle}
\newcommand{\eal}{\end{aligned}}
\newcommand{\torkance}{t}
\newcommand{\udot}{\overset{.}{u}}
\newcommand{\exponential}[1]{{\exp(#1)}}
\newcommand{\phandot}[1]{\overset{\phantom{.}}{#1}}
\newcommand{\phandag}{\phantom{\dagger}}
\newcommand{\Trace}{\text{Tr}}
\newcommand{\Bxc}{\Omega}
\setcounter{secnumdepth}{2}
\title{Spin-orbit torques in strained PtMnSb from first principles}
\author{Frank Freimuth$^{1,2}$}
\email[Corresp.~author:~]{f.freimuth@fz-juelich.de}
\author{Stefan Bl\"ugel$^{1}$}
\author{Yuriy Mokrousov$^{1,2}$}
\affiliation{$^1$Peter Gr\"unberg Institut and Institute for Advanced Simulation,
Forschungszentrum J\"ulich and JARA, 52425 J\"ulich, Germany}
\affiliation{$^2$Institute of Physics, Johannes Gutenberg University Mainz, 55099 Mainz, Germany
}
\date{\today}
\begin{abstract}
We compute spin-orbit torques (SOTs) in strained PtMnSb from first principles.
We consider both tetragonal strain and shear strain. We find a strong
linear
dependence of the field-like SOTs on these strains, while the
antidamping
SOT is only moderately sensitive to
shear strain and even
insensitive to tetragonal strain. 
We also study the dependence of the SOT on the magnetization
direction. In order to obtain analytical expressions suitable for
fitting
our numerical \textit{ab-initio} results we derive a general expansion
of the SOT in terms of all response tensors that are allowed by crystal symmetry.
Our expansion includes also higher-order terms beyond the usually
considered lowest order. We find that the dependence 
on the strain is much smaller for the higher-order terms than for the
lowest order terms. In order to judge the sensitivity of the SOT to
the
exchange correlation potential we compute the SOT in both GGA and LDA.
We find that the higher-order terms depend significantly on the
exchange-correlation potential, while the lowest order terms are
insensitive to it. Since the higher-order terms are small in
comparison
to the lowest order terms the total SOT is insensitive to the exchange
correlation potential in strained PtMnSb.
\end{abstract}
\maketitle
\section{Introduction}

The spin-orbit torque (SOT) allows us to switch
the magnetization by electric current in noncentrosymmetric
bulk crystals and in bilayers with structural inversion 
asymmetry~\cite{rmp_sot}. It therefore paves the way to novel
spintronic memory devices.
Among the noncentrosymmetric bulk crystals the
half-metallic half-Heusler compounds are promising for spintronics
applications~\cite{Galanakis_2006,Heusler_alloys_review,PhysRevB.95.024411,half_heuslers_novel_materials_energy_spintronics}. 
In particular, their high conduction-electron spin-polarization
 enhances for example the tunneling magnetoresistance and
the giant
magnetoresistance~\cite{,amr_cppgmr_nimnsb_multilayers,cppgmr_nimnsb_ag_spacers,nimnsb_based_MR,fully_epitaxial_cppgmr_ag_spacer},
and their half-metallicity suppresses the 
Gilbert damping~\cite{low_gilbert_half_metal}.

The SOT in the half-Heusler NiMnSb
depends strongly on the strain, which may be controlled by
varying the substrate~\cite{Ciccarelli_NiMnSb,2021unidirectional}.
Notably, NiMnSb thin films sputtered 
on GaAs substrates yield SOT effective fields per
applied current that are similar in magnitude 
to those in Pt/Co/AlO$_x$ magnetic bilayers~\cite{PhysRevMaterials.5.014413}.
Tetragonal strain adds Dresselhaus spin-orbit interaction (SOI) to the
microscopic half-Heusler Hamiltonian, while shear strain supplements
it with both 
Rashba and Dresselhaus SOI. 

The SOTs arising from Dresselhaus and Rashba SOI
correspond to the lowest order in the expansion of the SOT
with respect to the magnetization~\cite{PhysRevB.95.014403}. 
In magnetic bilayers the higher-order terms in this
expansion have been found to be sizeable
in experiments~\cite{symmetry_spin_orbit_torques}, and 
several theoretical works have therefore considered the
dependence of the SOT on the magnetization direction
in detail in these bilayer systems~\cite{PhysRevB.97.224426,anisotropic_sot_hanke,PhysRevMaterials.3.011401}.
However, in the case of half-Heusler crystals 
the higher-order contributions in the expansion of the
SOT in terms of the directional cosines of the magnetization 
have not yet been considered. Therefore, our symmetry analysis
of the SOT in this paper includes also the first higher-order terms
in the directional cosine expansion. Such angular expansions may
be used to fit experimental SOT
data~\cite{symmetry_spin_orbit_torques}.
In the present paper we use the angular expansion in order to fit
our \textit{ab-initio} data, which allows us to separate the SOT
into the lowest-order Dresselhaus and Rashba SOI contributions and
the remaining higher-order terms.

PtMnSb is  a promising material for spintronics applications.
Its half-metallicity has been established both experimentally and
theoretically.
It can be grown epitaxially on MgO(001)~\cite{pssr.201600439}
and on W(001)/MgO(001)~\cite{doi:10.1063/1.113262}.
It exhibits a giant magneto-optical Kerr
effect~\cite{doi:10.1063/1.93849,PtMnSb_oppeneer}, which makes it
attractive for magneto-optical recording. Furthermore, it exhibits
a negative anisotropic magnetoresistance and it
has been used for room-temperature
giant magnetoresistance devices~\cite{doi:10.1063/1.364925,Wen_2018}.
In this paper we discuss the SOT in
PtMnSb
with tetragonal and shear strain obtained from first principles
density-functional theory calculations.

This paper is structured as follows:
In Sec.~\ref{sec_symmetry} we discuss the form of the SOT expected
in half Heuslers based on the symmetry of the cubic, tetragonally
strained, and shear-strained crystals.
The tetragonally strained case is discussed in detail
in Sec.~\ref{sec_symmetry}, while the cubic and the
shear-strained cases are discussed in detail in the
Appendices~\ref{app_odd} and
\ref{app_even}.
In Sec.~\ref{sec_results} we present our \textit{ab-initio} results
on the SOTs in PtMnSb.
In Sec.~\ref{sec_formalism} we describe the computational details.
In Sec.~\ref{sec_results_odd_torque} we discuss the results on the odd
torque
and in Sec.~\ref{sec_results_even_torque} we discuss the results on the even
torque.
This paper ends with a summary in Sec.~\ref{sec_summary}.

\section{Symmetry of SOTs in half-Heusler crystals}
\label{sec_symmetry}
Similar to the conductivity tensor, which measures the response
of the electric current to an applied electric field in linear
response,
we introduce the torkance tensor to quantify the response of the
torque to an applied electric field~\cite{ibcsoit}.
The torque $\vn{T}$ acting on the magnetization in one crystal unit
cell 
is written as
\bege
\vn{T}=\sum_{ij}\hat{\vn{e}}_{i}
t_{ij}E_{j},
\ee
where $t_{ij}$
is the
torkance tensor,
 $E_{j}$ is the $j$-th component of the applied electric
field,
and $\hat{\vn{e}}_{i}$
is a unit vector in the $i$-th Cartesian direction.
In cubic and tetragonally strained PtMnSb
the crystal lattice vectors $\vn{a}$, $\vn{b}$, and $\vn{c}$ used in
the following sections are related to $\hat{\vn{e}}_{i}$
by $\vn{a}=a\hat{\vn{e}}_{1}$, $\vn{b}=b\hat{\vn{e}}_{2}$,
and $\vn{c}=c\hat{\vn{e}}_{3}$, where $a$, $b$, and $c$ are
the lattice constants.
In shear-strained PtMnSb
we choose the $\vn{a}$ and $\vn{b}$ axes as follows:
\bege
\begin{aligned}
&\vn{a}=a\left(\cos\frac{\epsilon}{2},\sin\frac{\epsilon}{2},0\right)^{\rm T}\\
&\vn{b}=a\left(\sin\frac{\epsilon}{2},\cos\frac{\epsilon}{2},0\right)^{\rm T},
\end{aligned}
\ee
where we use $\epsilon=90^{\circ}-\gamma$ to quantify the shear
strain,
and $\gamma$ is the angle between the $\vn{a}$ and $\vn{b}$ axes.

We separate the torkance into even and odd parts with respect to
inversion of the magnetization direction, 
i.e.,  $\vn{t}(\hat{\vn{M}})=\vn{t}^{\rm even}(\hat{\vn{M}})+\vn{t}^{\rm odd}(\hat{\vn{M}})$,
where $\vn{t}^{\rm even}(\hat{\vn{M}})=[\vn{t}(\hat{\vn{M}})+\vn{t}(-\hat{\vn{M}})]/2$ 
and $\vn{t}^{\rm odd}(\hat{\vn{M}})=[\vn{t}(\hat{\vn{M}})-\vn{t}(-\hat{\vn{M}})]/2$.
The corresponding even SOT is often referred to as the antidamping
SOT, while the odd SOT is often referred to as the field-like SOT.
$t_{ij}$ is an axial tensor of rank 2. It is possible to use
the symmetries of the half-Heusler crystal in order to determine
the form of  $t_{ij}$. In practice only the torque
perpendicular to the magnetization is of relevance and our 
\textit{ab-initio} approach described in Sec.~\ref{sec_formalism}
computes only this perpendicular component by construction.
However, in general, an
axial tensor of rank 2 consistent with the crystal symmetry may
predict
also a component of the torque that is parallel to the magnetization.
In order to avoid this irrelevant component we consider instead the
symmetry of
the
effective magnetic field $\vn{B}$ that one would have to apply
in order to generate a torque of the same size as the SOT.
After determining the symmetry-allowed form of the
response of $\vn{B}$ to an applied electric field we may 
subsequently obtain the torque 
from $\vn{T}=\mu \hat{\vn{M}} \times\vn{B} $.
Here, $\mu$ is the magnetic moment within one unit cell
and $\hat{\vn{M}}$ is its direction.
This approach guarantees that the torque $\vn{T}$ is perpendicular
to the magnetization such that it is not necessary to remove 
irrelevant contributions obtained from symmetry analysis.

\subsection{Odd torque}
The effective field of the odd torque can be expressed in terms
of the electric field $\vn{E}$ and the magnetization 
direction $\hat{\vn{M}}$ as follows:
\bege\label{eq_expand_odd}
B^{\rm{odd}}_{i}
=\chi^{({\rm a})}_{ij}E_{j}+\chi^{({\rm a})}_{ijkl}E_{j}\hat{M}_{k}\hat{M}_{l}+\dots
\ee
Here, $\chi^{({\rm a})}_{ij}$ is an axial tensor of second rank,
$\chi^{({\rm a})}_{ijkl}$ is an axial tensor of fourth rank and summation over
repeated indices is implied. Note that the effective field of the odd
torque
is even in the 
magnetization: $\vn{B}^{\rm odd}(\hat{\vn{M}})=\vn{B}^{\rm odd}(-\hat{\vn{M}})$
because of $\vn{T}^{\rm odd}=\mu \hat{\vn{M}} \times\vn{B}^{\rm odd}$ 
(in our notation the torque $\vn{T}$ carries 
the superscript 'odd', when it is odd in
the
magnetization, i.e., $\vn{T}^{\rm odd}(\hat{\vn{M}})=-\vn{T}^{\rm odd}(-\hat{\vn{M}})$,
while the effective magnetic field $\vn{B}$ carries 
the superscript 'odd', when it generates $\vn{T}^{\rm odd}$).
In order to express the symmetry-allowed 
tensors $\chi^{({\rm a})}_{ij}$ 
and $\chi^{({\rm a})}_{ijkl}$
in terms of
basis tensors we introduce the following notation to define these
basis tensors:
\bege
\delta^{(mn)}_{ij}=\delta_{im}\delta_{jn}\rightarrow\langle mn \rangle
\ee
and
\bege\label{eq_notation_tensors_4}
\delta^{(mnop)}_{ijkl}=\delta_{im}\delta_{jn}\delta_{ko}\delta_{lp}\rightarrow
\langle mnop \rangle.
\ee
The superscripts $(mn)$ and $(mnop)$ serve to label the basis tensors. As a
simple
example to illustrate the use of these basis tensors consider the unit
matrix. The unit matrix can be expressed as follows:
\bege
\delta_{ij}=\delta^{(11)}_{ij}+\delta^{(22)}_{ij}+\delta^{(33)}_{ij},
\ee
or simply $\langle 11\rangle+\langle 22\rangle+\langle 33\rangle$.
Similarly, any given tensor may be expressed in terms of these
basis tensors.
The symmetry-allowed form of the torkance tensor depends on the
crystallographic point group~\cite{PhysRevB.95.014403,birss}.
Cubic, tetragonally-strained, and shear-strained PtMnSb possess
different crystallographic point groups. Therefore, we discuss the
symmetry-allowed form of the torkance tensor separately for these
three cases in the following.
Note that in Eq.~\eqref{eq_expand_odd}
we expand the effective field only in terms of the applied electric field and in terms of
the magnetization but we do not expand it in terms of the strain.
This is a major difference to the treatment of e.g.\ the piezomagnetic
effects
in Ref.~\cite{birss},
where the strain itself is considered as a perturbation. 
Instead, we
assume here that the strain is constant and that it determines the symmetry-allowed
form of the response tensor by affecting the crystallographic point group.

\begin{threeparttable}
\caption{
List of axial tensors of ranks 2 and 4 allowed by symmetry in
tetragonally strained half Heuslers.
The notation introduced in Eq.~\eqref{eq_notation_tensors_4} is used.
Arrows indicate tensors that may be replaced by others due to
permutations
of indices, while \eqref{eq_a3_a9_a11_a1} denotes tensors
that may be replaced by others due to Eq.~\eqref{eq_a3_a9_a11_a1}.
}
\label{tab_tensors_odd_tetragonal}
\begin{ruledtabular}
\begin{tabular}{c|c||c|c|c|}
\#
&$\chi^{(\rm a\#)}$
&\#
&$\chi^{(\rm a\#)}$&Remark
\\
\hline
1 & $\langle22\rangle-\langle11\rangle$ 
&7 &$\langle3113\rangle-\langle3223\rangle$ &$\rightarrow\chi^{(\rm a5)}$\\
\hline
2 &$\langle2112\rangle-\langle1221\rangle$ 
&8 &$\langle2323\rangle-\langle1313\rangle$ &$\rightarrow\chi^{(\rm a6)}$\\
\hline
3 & $\langle1122\rangle-\langle2211\rangle$ 
&9 &$\langle2233\rangle-\langle1133\rangle$ &\\
\hline
4 &$\langle3322\rangle-\langle3311\rangle$ 
&10 &$\langle1212\rangle-\langle2121\rangle$ &$\rightarrow-\chi^{(\rm a2)}$\\
\hline
5 &$\langle3131\rangle-\langle3232\rangle$ 
&11 &$\langle2222\rangle-\langle1111\rangle$ &\eqref{eq_a3_a9_a11_a1}\\
\hline
6 & $\langle2332\rangle-\langle1331\rangle$ 
& & &\\
\hline
\end{tabular}
\end{ruledtabular}
\end{threeparttable}

\subsubsection{Tetragonal strain}
First, we consider the case of tetragonal strain. The cases of shear
strain and of cubic half Heuslers are discussed in Appendix~\ref{app_odd}.
For $a=b\ne c$ and $\alpha=\beta=\gamma=90^{\circ}$ (point group $\bar{4}2m$) we list the 11
axial tensors of rank 2 and 4 that are allowed by symmetry 
in Table~\ref{tab_tensors_odd_tetragonal}.
In Eq.~\eqref{eq_expand_odd} the indices $k$ 
and $l$ of $\chi^{(a)}_{ijkl}$ both couple to magnetization and are therefore
interchangeable.
Therefore, as indicated in Table~\ref{tab_tensors_odd_tetragonal} by arrows,
\bege\label{eq_reduce_number_of_tensors_odd}
\begin{aligned}
\chi^{({\rm a}10)}_{ijkl}&=-\chi^{({\rm a}2)}_{ijlk}\\
\chi^{({\rm a}7)}_{ijkl}&=\chi^{({\rm a}5)}_{ijlk}\\
\chi^{({\rm a}8)}_{ijkl}&=\chi^{({\rm a}6)}_{ijlk}.\\
\end{aligned}
\ee
Moreover, we find
\bege\label{eq_a3_a9_a11_a1}
(\chi^{({\rm a}3)}_{ijkl}-\chi^{({\rm a}9)}_{ijkl}-\chi^{({\rm a}11)}_{ijkl})\hat{M}_{k}\hat{M}_{l}=
-\chi^{({\rm a}1)}_{ij}.
\ee
Thus, we do not need to consider the tensors 10, 7, 8 and 11
% $\chi^{({\rm a}10)}_{ijkl}$, $\chi^{({\rm a}7)}_{ijkl}$ and
% $\chi^{({\rm a}8)}_{ijkl}$ $\chi^{({\rm a}11)}_{ijkl}$
when we express $\chi^{({\rm a})}_{ijkl}$ in terms of the
tensors in Table~\ref{tab_tensors_odd_tetragonal}.
Consequently, we can express the tensors in Eq.~\eqref{eq_expand_odd}
as 
\bege\label{eq_expansions_of_tensors}
\begin{aligned}
\chi^{({\rm a})}_{ij}=&\alpha_{1}\chi^{({\rm a}1)}_{ij}\\
\chi^{({\rm a})}_{ijkl}=&\alpha_{2}\chi^{({\rm a}2)}_{ijkl}
+\alpha_{3}\chi^{({\rm a}3)}_{ijkl}
+\alpha_{4}\chi^{({\rm a}4)}_{ijkl}+\\
&+\alpha_{5}\chi^{({\rm a}5)}_{ijkl}
+\alpha_{6}\chi^{({\rm a}6)}_{ijkl}
+\alpha_{7}\chi^{({\rm a}9)}_{ijkl}
\end{aligned}
\ee
in terms of 7 expansion coefficients $\alpha_{1},\dots,\alpha_{7}$.
The tensor $\chi^{({\rm a}1)}_{ij}=\delta^{(22)}_{ij}-\delta^{(11)}_{ij}$ describes the
effective SOT field from Dresselhaus SOI~\cite{2021unidirectional,PhysRevB.95.014403}. 
The tensors 2, 3, 4, 5, 6, and 9 describe higher-order contributions to the SOT,
which have not yet been discussed in the literature.

The odd torque $\vn{T}^{\rm odd}$ is related to its effective field 
by
\bege\label{eq_relate_todd_Bodd}
T_{i}^{\rm odd}
=
\Xi_{ij}
B_{j}^{\rm odd},
\ee
where
\bege\label{eq_cross_op}
\vn{\Xi}=
\mu
\begin{pmatrix}
0 &-\hat{M}_3 &\hat{M}_2\\
\hat{M}_3 &0 &-\hat{M}_1\\
-\hat{M}_2 &\hat{M}_1 &0
\end{pmatrix}.
\ee
Using Eq.~\eqref{eq_expand_odd},
Eq.~\eqref{eq_expansions_of_tensors},
and Eq.~\eqref{eq_relate_todd_Bodd}
we obtain
\bege
\vn{T}^{\rm odd}=\vn{t}^{\rm odd}\vn{E},
\ee
where
\bege\label{eq_expansions_of_tensors2}
t^{\rm odd}_{ij}=\mu\sum_{k=1}^{7}\alpha_{k}\vartheta^{({\rm odd} k)}_{ij}
\ee
with
\bege
\begin{aligned}
\vartheta^{(\rm{odd}1)}=&
\begin{pmatrix}
0 &-\hat{M}_3 &0\\
-\hat{M}_3 &0 &0\\
\hat{M}_2 &\hat{M}_1 &0
\end{pmatrix}\\
\vartheta^{(\rm{odd}2)}=&
\begin{pmatrix}
-\hat{M}_1\hat{M}_2\hat{M}_3 &0 &0\\
0 &-\hat{M}_1\hat{M}_2\hat{M}_3 &0\\
\hat{M}_1^2\hat{M}_2 &\hat{M}_1\hat{M}_2^2 &0
\end{pmatrix}
\\
\vartheta^{(\rm{odd}3)}=&
\begin{pmatrix}
0 &\hat{M}_1^2\hat{M}_3 &0\\
\hat{M}_2^2\hat{M}_3 &0 &0\\
-\hat{M}_2^3 &-\hat{M}_1^3 &0
\end{pmatrix}
\\
\vartheta^{(\rm{odd}4)}=&
\begin{pmatrix}
0 &0 &\hat{M}_2^3-\hat{M}_1^2\hat{M}_2\\
0 &0 &\hat{M}_1^3-\hat{M}_1\hat{M}_2^2\\
0 &0 &0
\end{pmatrix}
\\
\vartheta^{(\rm{odd}5)}=&
\begin{pmatrix}
\hat{M}_1\hat{M}_2\hat{M}_3 &-\hat{M}_2^2\hat{M}_3 &0\\
-\hat{M}_1^2\hat{M}_3 &\hat{M}_1\hat{M}_2\hat{M}_3 &0\\
0 &0 &0
\end{pmatrix}
\\
\vartheta^{(\rm{odd}6)}=&
\begin{pmatrix}
0 &0 &-\hat{M}_2\hat{M}_3^2\\
0 &0 &-\hat{M}_1\hat{M}_3^2\\
0 &0 &2\hat{M}_1\hat{M}_2\hat{M}_3
\end{pmatrix}
\\
\vartheta^{(\rm{odd}7)}=&
\begin{pmatrix}
0 &-\hat{M}_3^3 &0\\
-\hat{M}_3^3 &0 &0\\
\hat{M}_3^2\hat{M}_2 &\hat{M}_1\hat{M}_3^2 &0
\end{pmatrix}.
\\
\end{aligned}
\ee

Since
\bege\label{eq_linear_dependence_first_and_higher_orders}
\vartheta^{(\rm{odd}2)}+\vartheta^{(\rm{odd}5)}-\vartheta^{(\rm{odd}3)}+
\vartheta^{(\rm{odd}7)}=\vartheta^{(\rm{odd}1)}
\ee
we can set $\alpha_{7}=0$ 
in Eq.~\eqref{eq_expansions_of_tensors2}.
Thus, the odd torkance tensor can be expressed in terms of 6
tensors $\vartheta^{(\rm{odd}1)},...,\vartheta^{(\rm{odd}6)}$:
\bege\label{eq_final_fit_formula_odd}
t^{\rm odd}_{ij}=\sum_{k=1}^{6}\beta_{k}\vartheta^{({\rm odd} k)}_{ij}
\ee
with expansion coefficients $\beta_{1},...,\beta_{6}$.
By fitting Eq.~\eqref{eq_final_fit_formula_odd} to the odd torque
given for a set of magnetization directions, one may determine the coefficients
$\beta_{i}$ and subsequently use Eq.~\eqref{eq_final_fit_formula_odd}
to predict the odd torque for any magnetization direction.

\subsection{Even torque}
The effective field of the even torque can be expressed in terms
of the electric field $\vn{E}$ and the magnetization 
direction $\hat{\vn{M}}$ as follows:
\bege\label{eq_expand_even}
B^{\rm{even}}_{i}
=\chi^{({\rm p})}_{ijk}E_{j}\hat{M}_{k}
+\chi^{({\rm p})}_{ijklm}E_{j}
\hat{M}_{k}\hat{M}_{l}\hat{M}_{m}
+\dots
\ee
Here, $\chi^{({\rm p})}_{ijk}$ is a polar tensor of third rank,
$\chi^{({\rm p})}_{ijklm}$ is a polar tensor of fifth rank and summation over
repeated indices is implied. Note that the effective field of the even
torque
is odd in the magnetization.

\subsubsection{Tetragonal strain}
\begin{threeparttable}
\caption{
List of polar tensors of rank 3 and 5 allowed by symmetry in
tetragonally strained half Heuslers.
The notation introduced in Eq.~\eqref{eq_notation_tensors_4} is used.
Arrows indicate tensors that may be replaced by others due to
permutation of indices, while \eqref{eq_even_tetragonal_lindep}
denotes 
tensors that may be replaced
by others due to Eq.~\eqref{eq_even_tetragonal_lindep}.
}
\label{tab_tensors_even_tetr}
\begin{ruledtabular}
\begin{tabular}{c|c|c||c|c|c|}
\#
&$\chi^{(\rm p\#)}$&Note
&\#
&$\chi^{(\rm p\#)}$&Note
\\
\hline
1 & $\langle321\rangle+\langle312\rangle$ &
&18 &$\langle13121\rangle+\langle23212\rangle$ &$\rightarrow 17$\\
\hline
2 &$\langle231\rangle+\langle132\rangle$ &
&19 &$\langle22321\rangle+\langle11312\rangle$ &$\rightarrow 6$\\
\hline
3 & $\langle213\rangle+\langle123\rangle$ & 
&20 &$\langle11321\rangle+\langle22312\rangle$ &$\rightarrow 6$\\
\hline
4 &$\langle33231\rangle+\langle33132\rangle$ & 
&21 &$\langle31211\rangle+\langle32122\rangle$ &$\rightarrow 15$\\
\hline
5 &$\langle33321\rangle+\langle33312\rangle$ &$\rightarrow 4$
&22 &$\langle13211\rangle+\langle23122\rangle$ &$\rightarrow 17$\\
\hline
6 & $\langle22231\rangle+\langle11132\rangle$ &
&23 &$\langle32111\rangle+\langle31222\rangle$ &\eqref{eq_even_tetragonal_lindep}\\
\hline
7 &$\langle32331\rangle+\langle31332\rangle$ &
&24 &$\langle23111\rangle+\langle13222\rangle$ &\eqref{eq_even_tetragonal_lindep}\\
\hline
8 & $\langle23331\rangle+\langle13332\rangle$ & 
&25 &$\langle12311\rangle+\langle21322\rangle$&\eqref{eq_even_tetragonal_lindep} \\
\hline
9 &$\langle33213\rangle+\langle33123\rangle$ &$\rightarrow 4$ 
&26 &$\langle21311\rangle+\langle12322\rangle$&\eqref{eq_even_tetragonal_lindep}  \\
\hline
10 &$\langle32313\rangle+\langle31323\rangle$ &$\rightarrow 7$
&27 &$\langle11231\rangle+\langle22132\rangle$ &$\rightarrow 6$\\
\hline
11 & $\langle23313\rangle+\langle13323\rangle$ &$\rightarrow 8$
&28 &$\langle12131\rangle+\langle21232\rangle$& $\rightarrow 25$\\
\hline
12 &$\langle32133\rangle+\langle31233\rangle$ &$\rightarrow 7$
&29 &$\langle21131\rangle+\langle12232\rangle$ &$\rightarrow 26$ \\
\hline
13 & $\langle23133\rangle+\langle13233\rangle$ &$\rightarrow 8$
&30 &$\langle22213\rangle+\langle11123\rangle$&$\rightarrow 6$\\
\hline
14 &$\langle21333\rangle+\langle12333\rangle$ &
&31 &$\langle11213\rangle+\langle22123\rangle$ & $\rightarrow 6$\\
\hline
15 &$\langle31121\rangle+\langle32212\rangle$ &
&32 &$\langle12113\rangle+\langle21223\rangle$ &$\rightarrow 25$ \\
\hline
16 &$\langle32221\rangle+\langle31112\rangle$ &$\rightarrow 15$
&33 &$\langle21113\rangle+\langle12223\rangle$ &$\rightarrow 26$ \\
\hline
17 &$\langle23221\rangle+\langle13112\rangle$ &\eqref{eq_even_tetragonal_lindep}
& & &\\
\end{tabular}
\end{ruledtabular}
\end{threeparttable}

Here, we discuss the case of tetragonal strain. The cases of shear
strain and of cubic half Heuslers are discussed in Appendix~\ref{app_even}. 
For $a=b\ne c$ and $\alpha=\beta=\gamma=90^{\circ}$ we list the
polar tensors that are allowed by symmetry in Table~\ref{tab_tensors_even_tetr}.
In Eq.~\eqref{eq_expand_even} the indices $k$, $l$ and $m$
of $\chi^{(\rm p)}_{ijklm}$ are contracted with the magnetization
direction and are therefore interchangeable. Tensors that
are related to other tensors by interchange of the indices
$k$, $l$ and $m$ are specified in Table~\ref{tab_tensors_even_tetr}
by arrows.
These tensors do not need to be considered when we 
expand $\chi^{(\rm p)}_{ijklm}$. 
When considering the permutations of the indices $k$, $l$ and $m$
the list of  independent tensors that are
needed in the expansion of $\chi^{(\rm p)}_{ijk}$ and $\chi^{(\rm p)}_{ijklm}$
is therefore reduced to the following ones: 1, 2, 3, 4, 6, 7, 8, 14,
15, 17, 23, 24, 25, 26.

Due to the relations
\bege\label{eq_even_tetragonal_lindep}
\begin{aligned}
&\Xi_{ni}[
\chi^{({\rm p}4)}_{ijklm}
+
2\chi^{({\rm p}17)}_{ijklm}
]\hat{M}_k\hat{M}_l\hat{M}_m
=
0\\
&\Xi_{ni}[
\chi^{({\rm p}6)}_{ijklm}
+
\chi^{({\rm p}7)}_{ijklm}
+
\chi^{({\rm p}25)}_{ijklm}
]\hat{M}_k\hat{M}_l\hat{M}_m
=
0\\
&[
\chi^{({\rm p}7)}_{ijklm}
+
\chi^{({\rm p}15)}_{ijklm}
+
\chi^{({\rm p}23)}_{ijklm}
]\hat{M}_k\hat{M}_l\hat{M}_m
=
\chi^{({\rm p}1)}_{ijk}\hat{M}_k\\
&[
\chi^{({\rm p}8)}_{ijklm}
+
\chi^{({\rm p}17)}_{ijklm}
+
\chi^{({\rm p}24)}_{ijklm}
]\hat{M}_k\hat{M}_l\hat{M}_m
=
\chi^{({\rm p}2)}_{ijk}\hat{M}_k\\
&[
\chi^{({\rm p}14)}_{ijklm}
+
\chi^{({\rm p}25)}_{ijklm}
+
\chi^{({\rm p}26)}_{ijklm}
]\hat{M}_k\hat{M}_l\hat{M}_m
=
\chi^{({\rm p}3)}_{ijk}\hat{M}_k\\
\end{aligned}
\ee
we do not need to consider the tensors
17,
23,
24,
25,
and
26.
This leaves us with 3 polar tensors of rank 3
and 6 polar tensors of rank 5 to describe the
SOT effective magnetic field in the tetragonal
case, i.e., 9 tensors in total.

Using
$
\vn{T}^{\rm even}
=\mu
\hat{\vn{M}}
\times
\vn{B}^{\rm even},
$
and
$
\vn{T}^{\rm even}=\vn{t}^{\rm even}\vn{E},
$
we arrive at
\bege\label{eq_expansions_of_tensors2_even}
t^{\rm even}_{ij}=\mu\sum_{k=1}^{9}\gamma_{k}\vartheta^{({\rm even} k)}_{ij}
\ee
with
\bege
\begin{aligned}
\vartheta^{(\rm{even}1)}=&
\begin{pmatrix}
\hat{M}_2^2 &\hat{M}_1\hat{M}_2 &0\\
-\hat{M}_1\hat{M}_2 &-\hat{M}_1^2 &0\\
0 &0 &0
\end{pmatrix}\\
\vartheta^{(\rm{even}2)}=&
\begin{pmatrix}
0 &0 &-\hat{M}_1\hat{M}_3\\
0 &0 &\hat{M}_2\hat{M}_3\\
0 &0 &-\hat{M}_2^2+\hat{M}_1^2
\end{pmatrix}
\\
\vartheta^{(\rm{even}3)}=&
\begin{pmatrix}
-\hat{M}_3^2 &0 &0\\
0 &\hat{M}_3^2 &0\\
\hat{M}_1\hat{M}_3 &-\hat{M}_2\hat{M}_3 &0
\end{pmatrix}
\\
\vartheta^{(\rm{even}4)}=&
\begin{pmatrix}
0 &0 &2\hat{M}_1\hat{M}_2^2\hat{M}_3\\
0 &0 &-2\hat{M}_1^2\hat{M}_2\hat{M}_3\\
0 &0 &0
\end{pmatrix}
\\
\vartheta^{(\rm{even}5)}=&
\begin{pmatrix}
0 &-\hat{M}_1\hat{M}_2\hat{M}_3^2 &0\\
\hat{M}_1\hat{M}_2\hat{M}_3^2 &0 &0\\
-\hat{M}_1\hat{M}_2^2\hat{M}_3 &\hat{M}_1^2\hat{M}_2\hat{M}_3 &0
\end{pmatrix}
\\
\vartheta^{(\rm{even}6)}=&
\begin{pmatrix}
\hat{M}_2^2\hat{M}_3^2 &\hat{M}_1\hat{M}_2\hat{M}_3^2 &0\\
-\hat{M}_1\hat{M}_2\hat{M}_3^2 &-\hat{M}_1^2\hat{M}_3^2 &0\\
0 &0 &0
\end{pmatrix}
\\
\vartheta^{(\rm{even}7)}=&
\begin{pmatrix}
0 &0 &-\hat{M}_3^3\hat{M}_1\\
0 &0 &\hat{M}_3^3\hat{M}_2\\
0 &0 &\hat{M}_1^2\hat{M}_3^2-\hat{M}_2^2\hat{M}_3^2
\end{pmatrix}
\\
\vartheta^{(\rm{even}8)}=&
\begin{pmatrix}
-\hat{M}_3^4 &0 &0\\
0 &\hat{M}_3^4 &0\\
\hat{M}_3^3\hat{M}_1 &-\hat{M}_3^3\hat{M}_2 &0
\end{pmatrix}
\\
\vartheta^{(\rm{even}9)}=&
\begin{pmatrix}
\hat{M}_1^2\hat{M}_2^2 &\hat{M}_2^3\hat{M}_1 &0\\
-\hat{M}_1^3\hat{M}_2 &-\hat{M}_1^2\hat{M}_2^2 &0\\
0 &0 &0
\end{pmatrix}.
\\
\end{aligned}
\ee

\section{Results}
\label{sec_results}
\subsection{Computational details}
\label{sec_formalism}
\begin{figure}
\includegraphics[width=\linewidth,trim=12cm 21.5cm 1cm 2cm,clip]{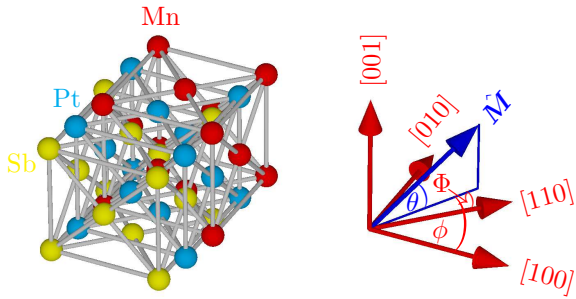}
\caption{\label{unitcellfigure}
Conventional unit cell of PtMnSb. The Pt, Mn, and Sb atoms each form
an fcc lattice individually. The polar and azimuthal angles of the
magnetization
are denoted by $\theta$ and $\phi$, respectively.
By $\Phi$ we denote the angle $\Phi=\phi-45^{\circ}$.
}
\end{figure}

We performed electronic structure calculations of 
PtMnSb based on the generalized gradient 
approximation (GGA)~\cite{PerdewBurkeEnzerhof} as implemented
in the FLEUR program~\cite{fleurcode}. The unit cell is shown
in Fig.~\ref{unitcellfigure}.
In order to judge how sensitive the SOT is to the
exchange correlation functional, we performed additional
calculations within the local density approximation
(LDA)~\cite{Barth_1972}.
We included SOI selfconsistently using the second
variation method~\cite{PhysRevB.42.5433}.
To compute cubic PtMnSb we use the experimental
lattice 
constant $a_{\rm cub}=c_{\rm cub}=11.72a_{0}$~\cite{doi:10.1063/1.113262}
in our calculations, where $a_{0}$ is Bohr's radius.
We considered 4 systems with different tetragonal 
strains $\eta=(c-c_{\rm cub})/c_{\rm cub}$:
$\eta=1.45\%$ ($a=11.49a_{0}$ and $c=11.89a_{0}$),
$\eta=0.723\%$ ($a=11.61a_{0}$ and $c=11.81a_{0}$),
$\eta=-0.723\%$ ($a=11.84a_{0}$ and $c=11.64a_{0}$), and
$\eta=-1.45\%$ ($a=11.95a_{0}$ and $c= 11.55a_{0}$).
Additionally, we considered 7 systems with different shear 
strains $\epsilon=\gamma-90^{\circ}$, where $\gamma$
is the angle between the $a$ axis and the $b$ axis:
$\epsilon=2^{\circ}$, $\epsilon=1^{\circ}$, $\epsilon=0.5^{\circ}$, $\epsilon=0.2^{\circ}$, $\epsilon=0.1^{\circ}$, $\epsilon=-0.1^{\circ}$, $\epsilon=-0.2^{\circ}$

After obtaining the electronic structure self-consistently
we generated maximally localized Wannier functions (MLWFs)
using the Wannier90 code~\cite{wannier90communitycode}
in order to calculate the SOTs according to the method described in Ref.~\cite{ibcsoit}
with the help of Wannier interpolation for computational speed-up.
We disentangled 44 MLWFs from 66 bands. For Mn and Pt
we used $sp^3d^2$, $d_{xy}$, $d_{yz}$ and $d_{zx}$ trial orbitals.
For Sb we employed $s$ and $p$ trial orbitals. 

The even torkance is given by ~\cite{ibcsoit}
\bege
\begin{aligned}
\label{eq_even_torque_constant_gamma}
t^{\rm even}_{ij}=&
\frac{e\hbar}{2\pi\mathcal{N}}
\sum_{\vn{k}n\ne m}
{\rm Im}
\left[
\langle
\psi^{\phantom{R}}_{\vn{k}n}
|
\mathcal{T}_{i}
|
\psi^{\phantom{R}}_{\vn{k}m}
\rangle
\langle
\psi^{\phantom{R}}_{\vn{k}m}
|
v_{j}
|
\psi^{\phantom{R}}_{\vn{k}n}
\rangle
\right]\Biggl\{\\
&\frac{\Gamma
(\mathcal{E}^{\phantom{R}}_{\vn{k}m}
-
\mathcal{E}^{\phantom{R}}_{\vn{k}n})
}{
\left[(\mathcal{E}^{\phantom{R}}_{\rm F}
-
\mathcal{E}^{\phantom{R}}_{\vn{k}n})^2+\Gamma^2\right]
\left[(\mathcal{E}^{\phantom{R}}_{\rm F}
-
\mathcal{E}^{\phantom{R}}_{\vn{k}m})^2+\Gamma^2\right]
}+\\
+&
\frac{
2\Gamma
}
{
\left[
\mathcal{E}^{\phantom{R}}_{\vn{k}n}
-
\mathcal{E}^{\phantom{R}}_{\vn{k}m}
\right]
\left[(\mathcal{E}^{\phantom{R}}_{\rm F}
-
\mathcal{E}^{\phantom{R}}_{\vn{k}m})^2+\Gamma^2\right]
}+\\
+&
\frac{
2
}
{
\left[
\mathcal{E}^{\phantom{R}}_{\vn{k}n}
-
\mathcal{E}^{\phantom{R}}_{\vn{k}m}
\right]^2
}
{\rm Im}\log
\frac{
\mathcal{E}^{\phantom{R}}_{\vn{k}m}
-
\mathcal{E}^{\phantom{R}}_{\rm F}-i\Gamma
}
{
\mathcal{E}^{\phantom{R}}_{\vn{k}n}
-
\mathcal{E}^{\phantom{R}}_{\rm F}-i\Gamma
}\Biggl\}
\end{aligned}
\ee
and the odd torkance is given by
\bege
\label{eq_odd_torque_constant_gamma}
t^{\rm odd}_{ij}=
\frac{e\hbar}{\pi\mathcal{N}}
\sum_{\vn{k}nm}
\frac{\Gamma^2
{\rm Re}
\left[
\langle
\psi^{\phantom{R}}_{\vn{k}n}
|
\mathcal{T}_{i}
|
\psi^{\phantom{R}}_{\vn{k}m}
\rangle
\langle
\psi^{\phantom{R}}_{\vn{k}m}
|
v_{j}
|
\psi^{\phantom{R}}_{\vn{k}n}
\rangle\right]
}{
\left[(\mathcal{E}^{\phantom{R}}_{\rm F}-\mathcal{E}^{\phantom{R}}_{\vn{k}n})^2+\Gamma^2\right]
\left[(\mathcal{E}^{\phantom{R}}_{\rm F}-\mathcal{E}^{\phantom{R}}_{\vn{k}m})^2+\Gamma^2\right]
},
\ee
where $\mathcal{N}$ is the number of $\vn{k}$-points used
to sample the Brillouin zone, $e$ is the elementary positive charge,
$\mathcal{T}_{i}$ is the $i$-th Cartesian component of the torque operator,
$v_{j}$ is the $j$-th Cartesian component of the velocity operator,
$\Gamma$ is the quasiparticle broadening,
and
$\psi^{\phantom{R}}_{\vn{k}n}$ and $\mathcal{E}^{\phantom{R}}_{\vn{k}n}$
denote the Bloch function
for band $n$ at $k$-point $\vn{k}$ and the
corresponding band energy, respectively.
A constant broadening of $\Gamma=25$~meV was used in the calculations
unless noted otherwise.

Due to the half-metallicity the spin magnetic moment per unit cell takes the integer
value $\mu= 4\mu_{\rm B}$ when SOI is not included
in the calculations, 
where $\mu_{\rm B}$ is Bohr's magneton. 
When we compute the magnetic moments contained in muffin-tin spheres
around the atoms, we find that Mn contributes most to the
total magnetic moment. In detail the atomic magnetic moments (in units
of $\mu_{\rm B}$) obtained in GGA (LDA) are as follows:
3.91 (3.8) on Mn,
0.11 (0.14) on Pt, and
-0.072 (-0.047) on Sb.
In our calculations of SOT  we
include SOI and therefore the magnetic moment slightly deviates from
the
integer value  $\mu= 4\mu_{\rm B}$. This deviation depends on the
strain and on the magnetization direction,
but it is at most 1\% for the strains that we consider. Therefore,
 $\mu\approx 4\mu_{\rm B}$ is very well satisfied in all our calculations.
When we present our \textit{ab-initio} results we 
use $ea_{0}\approx 8.478\times 10^{-30}$Cm as
the unit of torkance.
A torkance of one $ea_{0}$ 
corresponds therefore to an effective magnetic field of $B=ea_{0}
E/\mu\approx 0.229\,\mu$T
when the applied electric field is $E=1\,$V/m.

In Ref.~\cite{ibcsoit} we have shown that the odd SOT is
proportional to $1/\Gamma$ in the limit $\Gamma\rightarrow 0$, 
while the even SOT is
independent of $\Gamma$ in this limit.
Therefore, it may be convenient to discuss the odd SOT per applied
electric current, because this ratio is independent of $\Gamma$
in the limit of $\Gamma\rightarrow 0$. The resistivity of cubic PtMnSb
at $\Gamma=25$~meV is given by $\rho_{xx}=17\mu\Omega$cm,
which we computed using the equations given in Ref.~\cite{ibcsoit}.
Consequently, an odd torkance of one $ea_{0}$ at $\Gamma=25$~meV
corresponds to an effective magnetic field per electric current-density ratio 
of $B/j=ea_{0}E/(\mu j)=ea_{0}\rho_{xx}/\mu\approx 3.89\times10^{-14}$~Tm$^2$/A.

\subsection{Odd torque}
\label{sec_results_odd_torque}

\begin{figure}[H]
\includegraphics*[width=\linewidth]{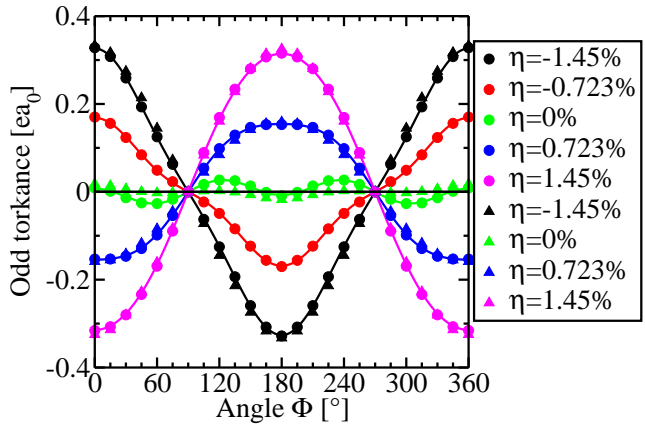}
\caption{\label{odd_torkance_ptmnsb_25meV_testfit}
Angular dependence of the odd torkance in 
PtMnSb for several strains $\eta=(c-c_{\rm cub})/c_{\rm cub}$
obtained in GGA (filled circles) and LDA (filled triangles)
when the electric current is applied along [110] direction
and when the magnetization is in-plane. $\Phi$ is the
angle between the magnetization and the [110] direction.
The component of the odd torque pointing in [001] direction is shown. 
Solid lines are fits to the GGA results
according to Eq.~\eqref{eq_final_fit_formula_odd}.
}
\end{figure}

In Fig.~\ref{odd_torkance_ptmnsb_25meV_testfit}
we show the odd torkance as a function of the azimuthal
angle of the magnetization for different tetragonal strains. 
Strain
increases the odd torkance significantly. 
At large strain the odd SOT is of the same order of magnitude 
as in experiments on NiMnSb~\cite{PhysRevMaterials.5.014413}.
A suitable substrate on which PtMnSb[100] grows under tetragonal strain
is W[100]. For W the theoretically estimated misfit strain is 2.1\%, while
the evaluation of diffraction data yields an estimated in-plane
tensile
strain of 0.31\%-0.52\%~\cite{doi:10.1063/1.113262}. 

In the tetragonal systems the differences between the
torkances computed with
GGA (filled circles) 
and LDA (filled triangles) are very small.
However, in the cubic system GGA and LDA differ even
qualitatively:
Here, the torkance has maxima close to 120$^{\circ}$ and
close to 240$^{\circ}$ when GGA is used. However, when LDA is used
it has a maximum at 0$^{\circ}$ instead.
When we use
Eq.~\eqref{eq_final_fit_formula_odd} to fit
the \textit{ab-initio} results we obtain very good agreement between
the fit and the data, as shown
in the figure. 

\begin{figure}[H]
\includegraphics*[width=\linewidth]{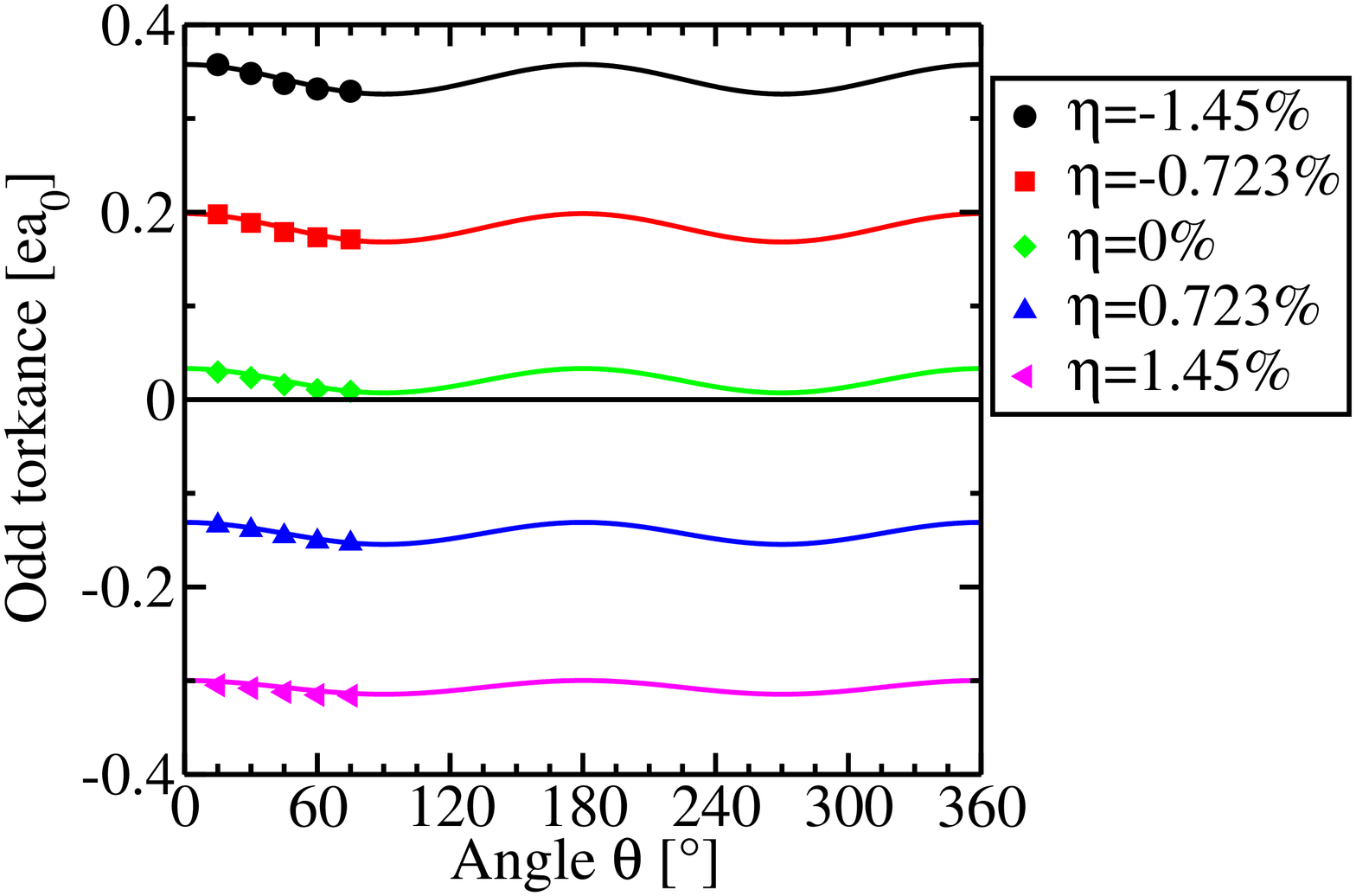}
\caption{\label{odd_torkance_ptmnsb_25meV_testfit_phi45theta}
Angular dependence of the odd torkance in 
PtMnSb for several strains $\eta=(c-c_{\rm cub})/c_{\rm cub}$ obtained
in GGA when
the electric current is applied in [110] direction. 
The magnetization is rotated from [001] direction ($\theta$=0)
to [110] direction ($\theta=90^{\circ}$). We show only the component
of the torque that is parallel to the unit vector $\vn{e}_{\theta}$ of
the spherical coordinate system.
\textit{Ab-initio} data are shown by symbols, while solid lines are fits
according to Eq.~\eqref{eq_final_fit_formula_odd}.}
\end{figure}

In Fig.~\ref{odd_torkance_ptmnsb_25meV_testfit_phi45theta}
we show the odd torkance as a function of the polar angle $\theta$.
It varies only moderately with the angle $\theta$, in contrast to the
strong variation with the angle $\phi$ shown 
in Fig.~\ref{odd_torkance_ptmnsb_25meV_testfit}. 
When the tetragonal strain is $\eta=$1.45\% the odd torkance is of the same
order
of magnitude as the even and odd torkances in magnetic bilayers
such as Co/Pt and Mn/W~\cite{ibcsoit}.

\begin{figure}[H]
\includegraphics*[width=\linewidth]{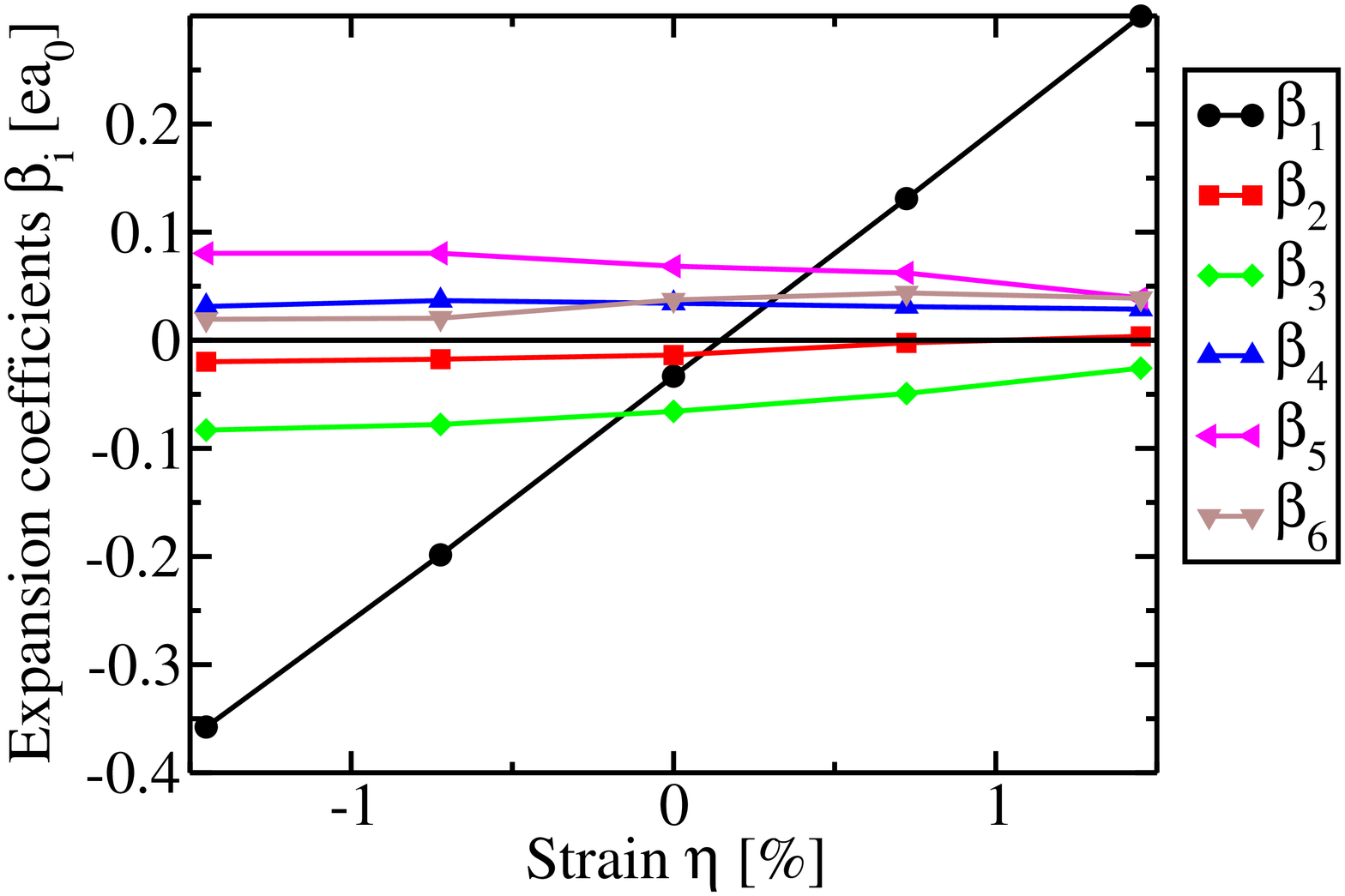}
\caption{\label{odd_expansion_vs_strain_25meV}
PtMnSb: Expansion 
coefficients $\beta_{i}$ in Eq.~\eqref{eq_final_fit_formula_odd}
for several strains $\eta=(c-c_{\rm cub})/c_{\rm cub}$
when the odd torkance is obtained
from GGA.
}
\end{figure}

In Fig.~\ref{odd_expansion_vs_strain_25meV} we show the
strain-dependence of the
parameters $\beta_{k}$ in Eq.~\eqref{eq_final_fit_formula_odd},
which we use to fit the \textit{ab-initio} results.
In cubic PtMnSb we find that 
the relations Eq.~\eqref{eq_cubic_tetragonal_relations} are 
satisfied well. 
The coefficient $\beta_{1}$ varies linearly with strain and depends
strongly on it, while the coefficients $\beta_{2}$ through $\beta_{6}$
are less sensitive to strain than $\beta_{1}$.
In Eq.~\eqref{eq_final_fit_formula_odd} $\beta_{1}$ is
the coefficient of $\vartheta^{({\rm odd} 1)}_{ij}$, which 
describes the SOT from Dresselhaus-type SOI.
However, $\beta_{1}$ does not vanish for zero strain.
Of course, this does not imply that there is a
Dresselhaus field at $\eta=0$. Instead, it is simply a manifestation of
Eq.~\eqref{eq_linear_dependence_first_and_higher_orders}, which shows
that $\vartheta^{({\rm odd} 1)}_{ij}$ is not linearly independent from
the higher order contributions described by the tensors 2, 3, 5, and
7.

\begin{figure}[H]
\includegraphics*[width=\linewidth]{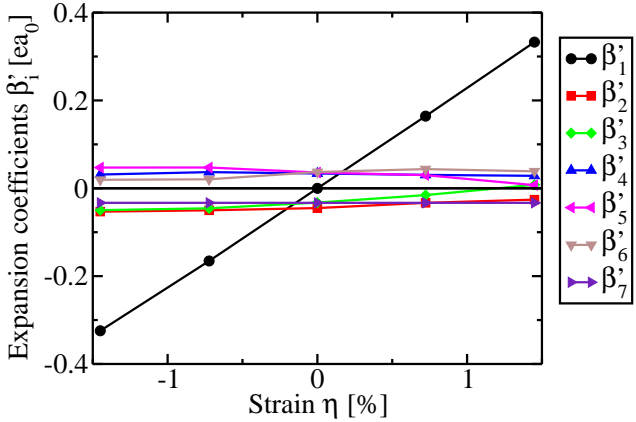}
\caption{\label{odd_expansion_vs_strain_25meV_beta7}
PtMnSb: Expansion 
coefficients $\beta^{'}_{i}$ in Eq.~\eqref{eq_final_fit_formula_odd2}
for several strains $\eta=(c-c_{\rm cub})/c_{\rm cub}$ when 
the odd torkance is obtained
from GGA.
}
\end{figure}

In Eq.~\eqref{eq_final_fit_formula_odd} we made the choice $\alpha_{7}$=0 
in order to get an unambiguous 
representation of the torque in terms of a set of
fitting parameters, which is only possible when we expand the torkance in terms of linearly
independent tensors.
However, it is possible to choose a different combination of tensors
such that the coefficient of $\vartheta^{({\rm odd} 1)}_{ij}$ is zero
at $\eta=0$. Such a combination has the advantage that one
may claim that the coefficient of $\vartheta^{({\rm odd} 1)}_{ij}$
corresponds to the Dresselhaus SOI.
For this purpose we perform a second fitting run after determining
the parameters $\beta_{1},...,\beta_{6}$ in
Eq.~\eqref{eq_final_fit_formula_odd}
in the first fitting run.
The second fitting run is based on
\bege\label{eq_final_fit_formula_odd2}
t^{\rm odd}_{ij}=\sum_{k=1}^{7}\beta'_{k}\vartheta^{({\rm odd}k)}_{ij},
\ee
where we fix $\beta'_{7}=\beta_{1}(\eta=0)$, while 
$\beta'_{1},...,\beta'_{6}$ are free fitting parameters.
As shown in Fig.~\ref{odd_expansion_vs_strain_25meV_beta7}
this two-step fitting procedure leads to $\beta'_{1}(\eta=0)=0$,
which can be understood easily from
Eq.~\eqref{eq_linear_dependence_first_and_higher_orders}.
We can thus claim that $\beta'_{1}$ describes the SOT from the 
Dresselhaus field.

\begin{figure}[H]
\includegraphics*[width=\linewidth]{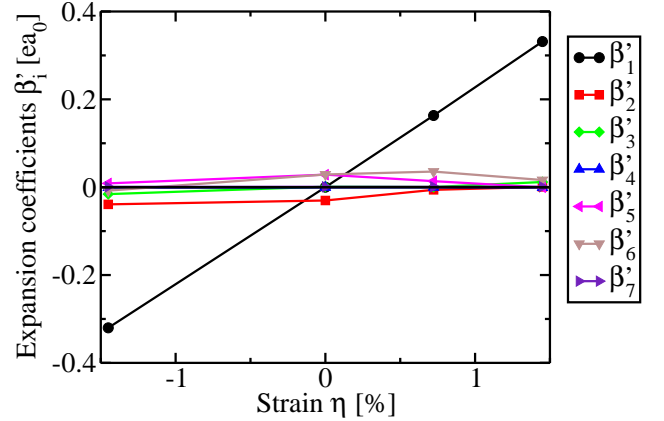}
\caption{\label{lda_odd_expansion_vs_strain_25meV_beta7}
PtMnSb: Expansion 
coefficients $\beta^{'}_{i}$ in Eq.~\eqref{eq_final_fit_formula_odd2}
for several strains $\eta=(c-c_{\rm cub})/c_{\rm cub}$ when
the odd torkance is obtained
from LDA.
}
\end{figure}

In Fig.~\ref{lda_odd_expansion_vs_strain_25meV_beta7}
we show the expansion coefficients for the SOT obtained from LDA.
Interestingly, the $\beta'_{1}$ 
in Fig.~\ref{odd_expansion_vs_strain_25meV_beta7} 
and Fig.~\ref{lda_odd_expansion_vs_strain_25meV_beta7}
differ by less than 1\%. Thus, the differences between
LDA and GGA, which are illustrated in Fig.~\ref{odd_torkance_ptmnsb_25meV_testfit}, 
are reflected
mostly by the differences in the higher-order coefficients
$\beta'_{2},...,\beta'_{7}$. These higher-order coefficients differ
significantly
between GGA and LDA. However, for sufficiently large strain the
contribution of the higher-order terms is relatively small compared
with
the Dresselhaus SOT described by $\beta'_{1}$. Therefore, the odd SOT
is insensitive to the choice of the exchange correlation potential in
strained
PtMnSb, as discussed already in Fig.~\ref{odd_torkance_ptmnsb_25meV_testfit}.

\begin{figure}[H]
\includegraphics*[width=\linewidth]{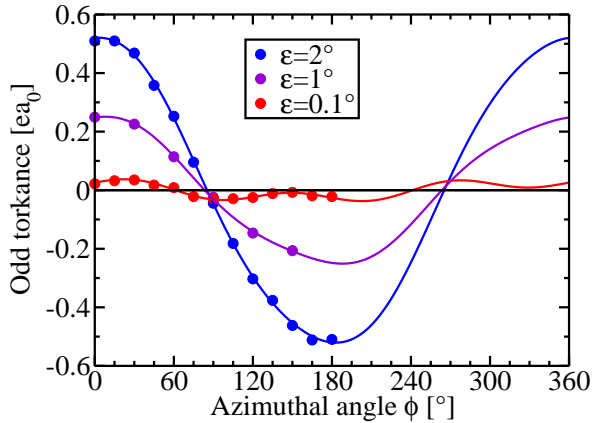}
\caption{\label{oddtorque_shearstrain_2deg_01deg}
Odd torkance in shear-strained PtMnSb obtained in GGA.
The current direction is along [100]. The magnetization is in-plane.
$\phi$ is the angle between the magnetization and the [100] direction.
\textit{Ab-initio} data are shown by filled circles, while solid lines are fits
according to Eq.~\eqref{eq_final_fit_formula_odd_shearstrain}.
}
\end{figure}

Next, we discuss the effect of shear strain $\epsilon$ on the odd torque.
In Fig.~\ref{oddtorque_shearstrain_2deg_01deg} 
we show the odd torkance as a function of the
azimuthal angle $\phi$ in the shear strained crystal.
The enhancement of the odd SOT with shear strain is similarly strong
as the enhancement with tetragonal strain.
When the shear strain is $\epsilon=2^{\circ}$ the odd torkance is of the same
order
of magnitude as the even and odd torkances in magnetic bilayers
such as Co/Pt and Mn/W~\cite{ibcsoit}.
The variation of the odd torkance with azimuthal angle $\phi$ is
similar
to the angular dependence in tetragonally strained PtMnSb shown in
Fig.~\ref{odd_torkance_ptmnsb_25meV_testfit}. 

\begin{figure}[H]
\includegraphics*[width=\linewidth]{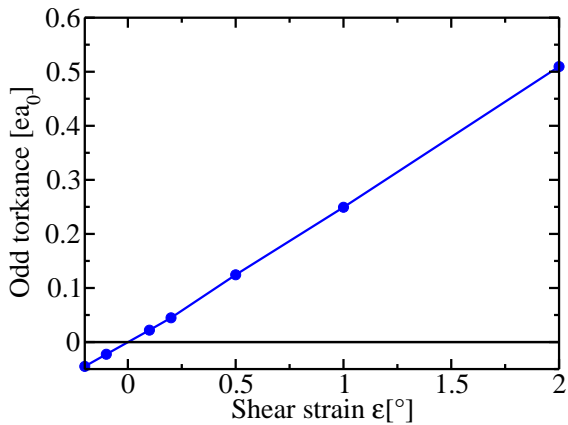}
\caption{\label{oddtorkance_vs_shearstrain}
Dependence of the odd torkance on shear-strain in PtMnSb when the
polar and azimuthal angles of magnetization 
are $\theta=90^{\circ}$ and $\phi=0^{\circ}$, 
respectively.
}
\end{figure}

At $\epsilon=2^{\circ}$ the odd torkance exhibits a maximum
at $\phi=0^{\circ}$. In order to investigate the dependence of this
maximum on $\epsilon$ we show in Fig.~\ref{oddtorkance_vs_shearstrain}
the odd torkance at $\phi=0^{\circ}$ as a function of $\epsilon$.
In the considered range the dependence on $\epsilon$ is
approximately linear.

\begin{figure}[H]
\includegraphics*[width=\linewidth]{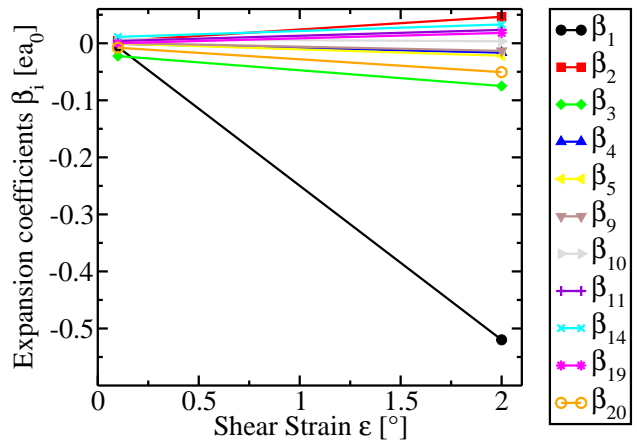}
\caption{\label{fitting_vs_shearstrain}
Expansion coefficients $\beta_{i}$
in Eq.~\eqref{eq_final_fit_formula_odd_shearstrain} of the
odd torkance in shear-strained PtMnSb obtained in GGA.
}
\end{figure}

In Fig.~\ref{fitting_vs_shearstrain} we show the expansion coefficients $\beta_{i}$
in Eq.~\eqref{eq_final_fit_formula_odd_shearstrain} of the
odd torkance. At large shear strain $\beta_{1}$ dominates clearly over
the other contributions, i.e., the SOT from Rashba SOI is dominant.
Since shear strain automatically implies tetragonal strain, a SOT from
Dresselhaus SOI -- described by $\beta_{2}$ -- is present as well, but
it is small compared to the SOT from Rashba SOI.

\subsection{Even Torque}
\label{sec_results_even_torque}

\begin{figure}[H]
\includegraphics*[width=\linewidth]{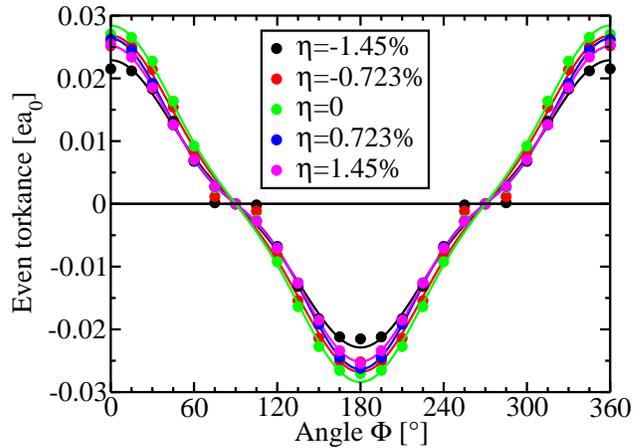}
\caption{\label{even_torkance_ptmnsb_25meV}
Angular dependence of the even torkance obtained within GGA in PtMnSb for several
tetragonal strains
$\eta=(c-c_{\rm cub})/c_{\rm cub}$ when the electric
current is applied along [110] direction and when the magnetization is
in-plane. 
$\Phi$ is the
angle between magnetization and the [110] direction. 
We show the component of the
even torque that is parallel to the unit vector $e_{\phi}$ of the
spherical coordinate system.
\textit{Ab-initio} data are shown by filled circles, while solid lines
are fits according to Eq.~\eqref{eq_expansions_of_tensors2_even}.
}
\end{figure}

In Fig.~\ref{even_torkance_ptmnsb_25meV}
we show the even torkance as a function of the azimuthal angle $\Phi$
for several tetragonal strains $\eta$. The even torque is considerably less
sensitive to strain than the odd torque. Additionally, at
$\eta=1.45\%$
the maximum even torkance is smaller than the maximum odd torkance by
a factor of 12.5. In contrast to magnetic bilayer 
systems such as Co/Pt~\cite{symmetry_spin_orbit_torques}, where the
even SOT is typically more important than the odd SOT, in PtMnSb the
odd SOT dominates.

\begin{figure}[H]
\includegraphics*[width=\linewidth]{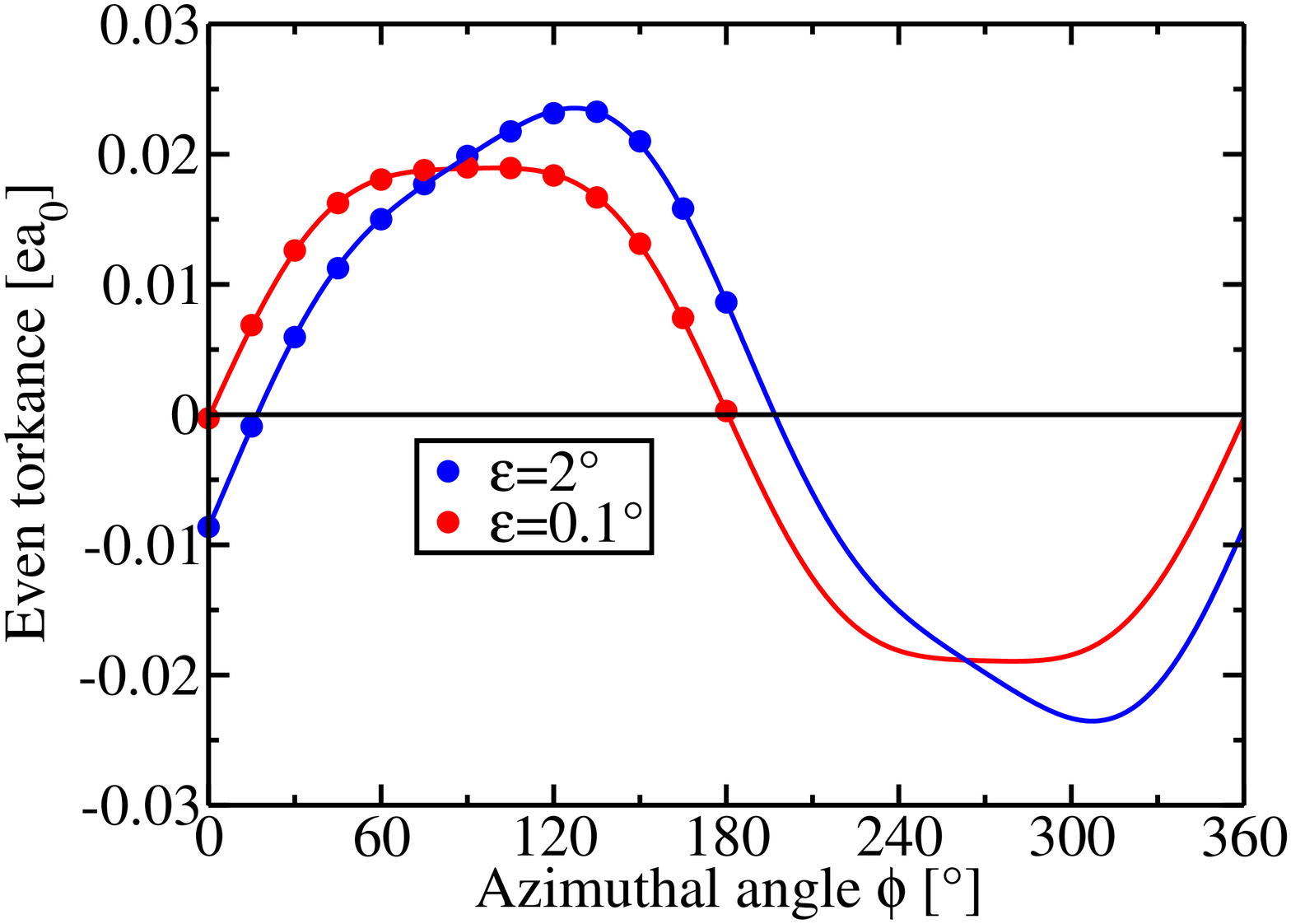}
\caption{\label{evensot_shearstrain_angular}
Angular dependence of the even torkance obtained within GGA in PtMnSb for several
shear strains
$\epsilon$ when the electric
current is applied along [100] direction and when the magnetization is in-plane. $\phi$ is the
angle between magnetization and the [100] direction. 
We show the component of the
even torque that is parallel to the unit vector $e_{\phi}$ of the
spherical coordinate system.
\textit{Ab-initio} data are shown by filled circles, while solid lines
are fits according to Eq.~\eqref{eq_even_shear_torkance_fit}.
}
\end{figure}

In Fig.~\ref{evensot_shearstrain_angular} we show the even torkance 
in shear-strained PtMnSb at $\Gamma=100$~meV. While the even torkance is more sensitive
to shear strain than to tetragonal strain, it is less sensitive to
shear stain than the odd torkance.

\section{Summary}
\label{sec_summary}
We discuss the constraints that crystal symmetry imposes on the
form of the SOT torkance tensor in half Heuslers with tetragonal or
shear strain. We discuss the lowest order tensors, which correspond
to Rashba and Dresselhaus SOI, but also higher order tensors.
We perform first principles DFT calculations of the SOT in half-Heusler
PtMnSb as a function of tetragonal and shear strain.
The odd torkance in PtMnSb depends strongly on tetragonal strain, which
we attribute to the Dresselhaus SOI. We find the SOT from Dresselhaus SOI 
to be insensitive to the exchange-correlation functional, i.e., the
differences
between GGA and LDA are negligible. In contrast, the higher-order
tensors differ substantially between GGA and LDA. However, these
higher-order
contributions are small in PtMnSb with tetragonal strain, such that
the
total odd torque in PtMnSb is insensitive to the exchange-correlation
functional.
The even torkance depends only weakly on tetragonal strain, but it
depends moderately strong on shear strain. The dependence of the
odd SOT on shear strain is similarly strong as its dependence on
tetragonal strain and it arises mostly from Rashba SOI.
In SOT-applications
PtMnSb should be grown on suitable substrates 
that maximize strain in order to obtain large torkances. 
Our results show that in strained PtMnSb torkances of the same order
of magnitude as in NiMnSb experiments may be achieved. 

\section*{Acknowledgments}We acknowledge financial support from
Leibniz Collaborative Excellence project OptiSPIN $-$ Optical Control
of Nanoscale Spin Textures, and funding  under SPP 2137 ``Skyrmionics"
of the DFG.
We gratefully acknowledge financial support from the European Research
Council (ERC) under
the European Union's Horizon 2020 research and innovation
program (Grant No. 856538, project ``3D MAGiC''). 
The work was also supported by the Deutsche Forschungsgemeinschaft
(DFG, German Research Foundation) $-$ TRR 173 $-$ 268565370 (project
A11), TRR 288 $-$ 422213477 (project B06).  We  also gratefully
acknowledge the J\"ulich
Supercomputing Centre and RWTH Aachen University for providing
computational
resources under project No. jiff40.

\appendix
\section{Symmetry analysis for the odd torque in cubic half
  Heuslers and in half Heuslers under shear strain}
\label{app_odd}

\subsubsection{Cubic PtMnSb}
In the following we discuss the odd torque for the case of cubic half Heuslers, 
i.e., $a=b=c$ and $\alpha=\beta=\gamma=90^{\circ}$ (point group $\bar{4}3m$). 
In contrast to the case with tetragonal strain, symmetry does not
allow
axial tensors of rank 2 in cubic PtMnSb~\cite{PhysRevB.95.014403}.
The following
axial tensors of rank 4 are allowed by symmetry:
\bege\label{eq_axial_tensors_fourth_rank_cubic}
\begin{aligned}
\chi^{({\rm a}12)}_{ijkl}=&
\delta^{(2121)}_{ijkl}-\delta^{(1212)}_{ijkl}
-\delta^{(3131)}_{ijkl}\\
&+\delta^{(3232)}_{ijkl}
+\delta^{(1313)}_{ijkl}-\delta^{(2323)}_{ijkl},
\\
\chi^{({\rm a}13)}_{ijkl}=&-\delta^{(2211)}_{ijkl}+\delta^{(1122)}_{ijkl}
+\delta^{(3311)}_{ijkl}\\
&-\delta^{(3322)}_{ijkl}
-\delta^{(1133)}_{ijkl}+\delta^{(2233)}_{ijkl},
\\
\chi^{({\rm a}14)}_{ijkl}=&-\delta^{(1221)}_{ijkl}+\delta^{(2112)}_{ijkl}
+\delta^{(1331)}_{ijkl}\\
&-\delta^{(2332)}_{ijkl}
-\delta^{(3113)}_{ijkl}+\delta^{(3223)}_{ijkl}.
\\
\end{aligned}
\ee
Since the indices $k$ and $l$ of $\chi_{ijkl}$ 
both couple to magnetization in Eq.~\eqref{eq_expand_odd} 
and since 
\bege
\chi^{({\rm a}12)}_{ijkl}\hat{M}_{k}\hat{M}_{l}
=
\chi^{({\rm a}14)}_{ijkl}\hat{M}_{k}\hat{M}_{l}
\ee
we do not need to consider $\chi^{({\rm a}14)}$ when 
we expand $\chi^{({\rm a})}_{ijkl}$ in terms of the tensors 
in Eq.~\eqref{eq_axial_tensors_fourth_rank_cubic}.
Comparison of these tensors to Table~\ref{tab_tensors_odd_tetragonal} yields
\bege\label{eq_axial_tensors_fourth_rank_cubic3}
\begin{aligned}
\chi^{({\rm a}12)}_{ijkl}=&
\chi^{({\rm a}2)}_{ijlk}
-\chi^{({\rm a}5)}_{ijkl}
-\chi^{({\rm a}6)}_{ijlk},\\
\chi^{({\rm a}13)}_{ijkl}=&
\chi^{({\rm a}3)}_{ijkl}
-\chi^{({\rm a}4)}_{ijkl}
+\chi^{({\rm a}9)}_{ijkl}.
\\
\end{aligned}
\ee
Thus, for the cubic half Heuslers, we can express the tensors in
Eq.~\eqref{eq_expand_odd} as follows:
\bege\label{eq_expansions_of_tensors_cubic}
\begin{aligned}
\chi^{({\rm a})}_{ij}=&0\\
\chi^{({\rm a})}_{ijkl}=&\alpha_{12}\chi^{({\rm a}12)}_{ijkl}
+\alpha_{13}\chi^{({\rm a}13)}_{ijkl},
\end{aligned}
\ee
with two coefficients $\alpha_{12}$ and
$\alpha_{13}$~\cite{PhysRevB.95.014403}.

The corresponding torkance is given by
\bege\label{eq_fit_odd_sot_cubic}
t^{\rm odd}_{ij}=
\mu
\sum_{k=12}^{13}
\alpha_{k}^{\phantom{(k)}}
\vartheta^{({\rm odd}k)}_{ij}
=\sum_{k=12}^{13}
\beta_{k}^{\phantom{(k)}}
\vartheta^{({\rm odd}k)}_{ij}
\ee
with
\bege
\begin{aligned}
&\vartheta^{({\rm odd}12)}=\vartheta^{({\rm odd}2)}
-\vartheta^{({\rm odd}5)}-\vartheta^{({\rm odd}6)}=\\
&\begin{pmatrix}
-2\hat{M}_{1}\hat{M}_{2}\hat{M}_{3} &\hat{M}_2^2\hat{M}_3 &\hat{M}_2\hat{M}_3^2\\
\hat{M}_1^2\hat{M}_3 &-2\hat{M}_1\hat{M}_2\hat{M}_3 &\hat{M}_1\hat{M}_3^2\\
\hat{M}_1^2\hat{M}_2 &\hat{M}_1\hat{M}_2^2 &-2\hat{M}_1\hat{M}_2\hat{M}_3
\end{pmatrix}\\
\end{aligned}
\ee
and
\bege
\begin{aligned}
&\vartheta^{({\rm odd}13)}=\vartheta^{({\rm odd}3)}
-\vartheta^{({\rm odd}4)}+\vartheta^{({\rm odd}7)}=\\
&\begin{pmatrix}
0 &\hat{M}_1^2\hat{M}_3-\hat{M}_3^3 &-\hat{M}_2^3+\hat{M}_1^2\hat{M}_2\\
\hat{M}_2^2\hat{M}_3-\hat{M}_3^3 &0 &-\hat{M}_1^3+\hat{M}_1\hat{M}_2^2\\
-\hat{M}_2^3+\hat{M}_3^2\hat{M}_2 &-\hat{M}_1^3+\hat{M}_1\hat{M}_3^2 &0
\end{pmatrix}.
\\
\end{aligned}
\ee

Instead of using Eq.~\eqref{eq_fit_odd_sot_cubic} to fit the
odd torkance in cubic PtMnSb one can of course also use 
Eq.~\eqref{eq_final_fit_formula_odd}. By equating
Eq.~\eqref{eq_fit_odd_sot_cubic} and Eq.~\eqref{eq_final_fit_formula_odd}
we find that in cubic PtMnSb the following relations should be satisfied:
\bege\label{eq_cubic_tetragonal_relations}
\begin{aligned}
\beta_{1}&=-\beta_{4},\\
2\beta_{1}&=\beta_{3},\\
2\beta_{1}&=-\beta_{2}-\beta_{5},\\
\beta_{6}&=\beta_{1}+\beta_{5}.\\
\end{aligned}
\ee

\subsubsection{Shear strain}

\begin{threeparttable}
\caption{
List of axial tensors of rank 2 and 4 allowed by symmetry in
shear-strained half Heuslers.
The notation introduced in Eq.~\eqref{eq_notation_tensors_4} is used.
Arrows indicate tensors that may be replaced by others due to 
permutation of indices, while \eqref{eq_shear_odd_linrel} denotes
tensors that may be replaced by others due to Eq.~\eqref{eq_shear_odd_linrel}.
}
\label{tab_tensors_odd_shear}
\begin{ruledtabular}
\begin{tabular}{c|c|c||c|c|c|}
\#
&$\chi^{(\rm a\#)}$&Note
&\#
&$\chi^{(\rm a\#)}$&Note
\\
\hline
1 & $\langle21\rangle-\langle12\rangle$ &
&12 &$\langle3132\rangle-\langle3231\rangle$ &\eqref{eq_shear_odd_linrel}\\
\hline
2 &$\langle22\rangle-\langle11\rangle$ &
&13 &$\langle3232\rangle-\langle3131\rangle$ &\eqref{eq_shear_odd_linrel}\\
\hline
3 & $\langle2121\rangle-\langle1212\rangle$ &
&14 &$\langle2332\rangle-\langle1331\rangle$ &\\
\hline
4 &$\langle2221\rangle-\langle1112\rangle$ &
&15 &$\langle2313\rangle-\langle1323\rangle$ &$\rightarrow 5$\\
\hline
5 &$\langle2331\rangle-\langle1332\rangle$ &
&16 &$\langle3123\rangle-\langle3213\rangle$ &$\rightarrow 12$\\
\hline
6 & $\langle2112\rangle-\langle1221\rangle$ &$\rightarrow 3$
&17 &$\langle3223\rangle-\langle3113\rangle$ &$\rightarrow 13$\\
\hline
7 &$\langle2212\rangle-\langle1121\rangle$ &$\rightarrow 4$
&18 &$\langle2323\rangle-\langle1313\rangle$ &$\rightarrow 14$\\
\hline
8 & $\langle3312\rangle-\langle3321\rangle$ &$\emptyset$
&19 &$\langle2133\rangle-\langle1233\rangle$& \\
\hline
9 &$\langle2122\rangle-\langle1211\rangle$ &
&20 &$\langle2233\rangle-\langle1133\rangle$&  \\
\hline
10 &$\langle2222\rangle-\langle1111\rangle$ &
&21 &$\langle2111\rangle-\langle1222\rangle$ &\eqref{eq_shear_odd_linrel}\\
\hline
11 & $\langle3322\rangle-\langle3311\rangle$ &
&22 &$\langle2211\rangle-\langle1122\rangle$&\eqref{eq_shear_odd_linrel}\\
\hline
\end{tabular}
\end{ruledtabular}
\end{threeparttable}

Finally, we discuss the odd torque in the presence of shear strain.

We present the axial tensors of rank 2 and rank 4 that are allowed by
symmetry in shear-strained half-Heuslers in 
Table~\ref{tab_tensors_odd_shear}.
As indicated by arrows in the Table, tensors 6, 7, 15, 16, 17, and 18 
do not need to be considered because both indices $k$ and $l$
of $\chi^{(a)}_{ijkl}$ couple to magnetization in
Eq.~\eqref{eq_expand_odd} and these tensors may therefore be replaced by others.
Additionally, tensor 8 does not need to be considered, because it
evaluates to zero when both indices $k$ and $l$
of $\chi^{(a)}_{ijkl}$ are contracted with the magnetization.
Tensor 1 describes the SOT effective field from
Rashba SOI, tensor 2 describes the SOT effective field from 
Dresselhaus SOI~\cite{PhysRevB.95.014403},
and the remaining tensors describe higher-order contributions that
have not yet been discussed in the literature. Tensor 2 appears also
in the case of tetragonal strain, see the first tensor in 
Table~\ref{tab_tensors_odd_tetragonal}. This is expected, because
shear strain is automatically accompanied by tetragonal strain.

The corresponding torkance may be written as
\bege \label{eq_final_fit_formula_odd_shearstrain}
t^{\rm odd}_{ij}=\sum_{\#=1}^{2}\beta_{\#}\Xi_{im}\chi^{(\rm
  a\#)}_{mj}+
\sum_{\#=3}^{22}\beta_{\#}\Xi_{im}
\chi^{({\rm a\#})}_{mjkl}\hat{M}_{k}\hat{M}_{l},
\ee
where the matrix $\vn{\Xi}$ is defined in
Eq.~\eqref{eq_cross_op}. As discussed above, one may
set $\beta_{\#}=0$ for all tensors $\#$ indicated by an arrow or by $\emptyset$
in Table~\ref{tab_tensors_odd_shear}, i.e., $\beta_6=0, \beta_7=0, \beta_8=0,
\beta_{15}=0,\dots$. However, due to the relations
\bege\label{eq_shear_odd_linrel}
\begin{aligned}
&\Xi_{im}\chi^{(\rm
  a1)}_{mj}=
\Xi_{im}
\chi^{({\rm a4})}_{mjkl}\hat{M}_{k}\hat{M}_{l}
-
\Xi_{im}
\chi^{({\rm a12})}_{mjkl}\hat{M}_{k}\hat{M}_{l}
+\\
&\quad+
\Xi_{im}
\chi^{({\rm a19})}_{mjkl}\hat{M}_{k}\hat{M}_{l}
+
\Xi_{im}
\chi^{({\rm a21})}_{mjkl}\hat{M}_{k}\hat{M}_{l},\\
&\Xi_{im}\chi^{(\rm
  a2)}_{mj}=
\Xi_{im}
\chi^{({\rm a10})}_{mjkl}\hat{M}_{k}\hat{M}_{l}
+
\Xi_{im}
\chi^{({\rm a20})}_{mjkl}\hat{M}_{k}\hat{M}_{l}
+\\
&\quad+
\Xi_{im}
\chi^{({\rm a22})}_{mjkl}\hat{M}_{k}\hat{M}_{l},\\
&\Xi_{im}\chi^{(\rm
  a3)}_{mj}=
\Xi_{im}
\chi^{({\rm a10})}_{mjkl}\hat{M}_{k}\hat{M}_{l}
+
\Xi_{im}
\chi^{({\rm a13})}_{mjkl}\hat{M}_{k}\hat{M}_{l},\\
&\Xi_{im}
\left[
\chi^{({\rm a4})}_{mjkl}
-
\Xi_{im}
\chi^{({\rm a9})}_{mjkl}
-
\Xi_{im}
\chi^{({\rm a12})}_{mjkl}
\right]
\hat{M}_{k}\hat{M}_{l}
=0
\end{aligned}
\ee
the remaining tensors in
Eq.~\eqref{eq_final_fit_formula_odd_shearstrain} are not linearly
independent.
Therefore, we may additionally choose 
$\beta_{21}=0$,
$\beta_{22}=0$,
$\beta_{13}=0$,
 and $\beta_{12}=0$.
Thus, only 11 independent tensors need to be considered in
Eq.~\eqref{eq_final_fit_formula_odd_shearstrain}
with 11 corresponding fitting parameters $\beta_{\#}$.

\section{Symmetry analysis for the even torque in cubic half
  Heuslers and in half Heuslers under shear strain}

\label{app_even}

\subsubsection{Cubic PtMnSb}
When
$a=b=c$ and $\alpha=\beta=\gamma=90^{\circ}$,
symmetry allows 11 polar tensors of rank 3 and rank 5,
which we list in Table~\ref{tab_tensors_even_cubic}.

\begin{threeparttable}
\caption{
List of polar tensors of rank 3 and 5 allowed by symmetry in cubic
half Heuslers.
The notation introduced in Eq.~\eqref{eq_notation_tensors_4} is used.
Arrows indicate tensors that may be replaced by others.
}
\label{tab_tensors_even_cubic}
\begin{ruledtabular}
\begin{tabular}{c|c|c|}
\#
&$\chi^{(\rm p\#)}$&Note
\\
\hline
1 & 
$\phantom{a}_{\langle321\rangle+\langle231\rangle
+\langle312\rangle+\langle132\rangle
+\langle213\rangle+\langle123\rangle}$
&\\
\hline
2 & 
$\!\!\!\!\phantom{a}_{-\langle13121\rangle-\langle12131\rangle
-\langle23212\rangle-\langle21232\rangle
-\langle32313\rangle-\langle31323\rangle}\!$
&\\
\hline
3 & 
$\phantom{a}_{\langle32221\rangle+\langle23331\rangle
+\langle31112\rangle+\langle13332\rangle
+\langle21113\rangle+\langle12223\rangle}\!$
&\\
\hline
4 & 
$\phantom{a}_{\langle31211\rangle+\langle21311\rangle
+\langle32122\rangle+\langle12322\rangle
+\langle23133\rangle+\langle13233\rangle}\!$
&$\rightarrow\!\!\chi^{(\rm p 3)}$\!\\
\hline
5 & 
$\phantom{a}_{\langle32111\rangle+\langle23111\rangle
+\langle31222\rangle+\langle13222\rangle
+\langle21333\rangle+\langle12333\rangle}\!$
&\\
\hline
6 & 
$\phantom{a}_{\langle13211\rangle+\langle12311\rangle
+\langle23122\rangle+\langle21322\rangle
+\langle32133\rangle+\langle31233\rangle}\!$
&$\rightarrow\!\!\chi^{(\rm p 2)}$\!\\
\hline
7 & 
$\phantom{a}_{\langle23221\rangle+\langle32331\rangle
+\langle13112\rangle+\langle31332\rangle
+\langle12113\rangle+\langle21223\rangle}\!$
&\\
\hline
8 & 
$\phantom{a}_{\langle11321\rangle+\langle11231\rangle
+\langle22312\rangle+\langle22132\rangle
+\langle33213\rangle+\langle33123\rangle}\!$
&\\
\hline
9 & 
$\!\!\!\!\phantom{a}_{-\langle33321\rangle-\langle22231\rangle
-\langle33312\rangle-\langle11132\rangle
-\langle22213\rangle-\langle11123\rangle}\!$
&$\rightarrow\!\!\chi^{(\rm p 8)}$\!\\
\hline
10 & 
$\!\!\!\!\phantom{a}_{-\langle31121\rangle-\langle21131\rangle
-\langle32212\rangle-\langle12232\rangle
-\langle23313\rangle-\langle13323\rangle}\!$
&\\
\hline
11 & 
$\!\!\!\!\phantom{a}_{-\langle22321\rangle-\langle33231\rangle
-\langle11312\rangle-\langle33132\rangle
-\langle11213\rangle-\langle22123\rangle}\!$
&\\
\hline
\end{tabular}
\end{ruledtabular}
\end{threeparttable}

In Eq.~\eqref{eq_expand_even} the last three indices are
contracted with the magnetization. Therefore, for the purpose
of application in Eq.~\eqref{eq_expand_even}, $\chi^{({\rm p}6)}_{ijklm}$
is equivalent with $\chi^{({\rm p}2)}_{ijklm}$,
$\chi^{({\rm p}4)}_{ijklm}$
is equivalent with $\chi^{({\rm p}3)}_{ijklm}$,
and $\chi^{({\rm p}9)}_{ijklm}$
is equivalent with $\chi^{({\rm p}8)}_{ijklm}$, as indicated in the Table.
Consequently, we do not need to consider $\chi^{({\rm p}6)}_{ijklm}$,
$\chi^{({\rm p}4)}_{ijklm}$, and $\chi^{({\rm p}9)}_{ijklm}$. Thus,
the contribution to the effective field of the even SOT which
is third order in $\hat{\vn{M}}$ can be expressed in terms of the tensor
\bege
\begin{aligned}
&\chi^{({\rm p})}_{ijklm}=
\alpha_{2}^{\phantom{1}}\chi^{({\rm p2})}_{ijklm}
+\alpha_{3}^{\phantom{1}}\chi^{({\rm p3})}_{ijklm}
+\alpha_{5}^{\phantom{1}}\chi^{({\rm p5})}_{ijklm}
+\alpha_{7}^{\phantom{1}}\chi^{({\rm p7})}_{ijklm}\\
&+\alpha_{8}^{\phantom{1}}\chi^{({\rm p8})}_{ijklm}
+\alpha_{10}^{\phantom{1}}\chi^{({\rm p10})}_{ijklm}
+\alpha_{11}^{\phantom{1}}\chi^{({\rm p11})}_{ijklm}
.
\end{aligned}
\ee

% The following third rank polar tensor is allowed by symmetry:
% \bege\label{eq_polar_tensors_third_rank_nostrain}
% \begin{aligned}
% \chi^{({\rm p}1)}_{ijk}=&\delta^{(321)}_{ijk}+\delta^{(231)}_{ijk}
% +\delta^{(312)}_{ijk}+\delta^{(132)}_{ijk}
% +\delta^{(213)}_{ijk}+\delta^{(123)}_{ijk}
% \\
% \end{aligned}
% \ee
% The effective field of the even SOT which is first order in $\vn{M}$
% is therefore given by
% \bege
% B_{i}^{\rm even}=\alpha^{\phantom{1}}_{1}
% \chi^{({\rm p}1)}_{ij} E_{j}^{\phantom{1}}
% \ee
% where $\chi^{({\rm p}1)}_{ij}(\vn{M})=\chi^{({\rm p}1)}_{ijk}M_{k}^{\phantom{1}}$, 
% or explicitly 
% \bege\label{eq_even_eff_field_nostr_mf}
% \chi^{({\rm p1})}(\vn{M})=
% \begin{pmatrix}
% 0 &M_z  & M_y\\
% M_z  &0 &M_x \\
% M_y &M_x &0
% \end{pmatrix}
% \ee

\subsubsection{Shear strain}
Finally, we consider the case of shear strain.
In Table~\ref{tab_tensors_even_shear}
we present the polar tensors of rank 3 and 5 allowed by
symmetry in shear-strained half Heuslers.
As indicated in the Table by arrows, several tensors
may be replaced by others, because in Eq.~\eqref{eq_expand_even} 
the indices $k$, $l$ and $m$
of $\chi^{(\rm p)}_{ijklm}$ are interchangeable.
\begin{threeparttable}
\caption{
List of polar tensors of rank 3 and 5 allowed by symmetry in
shear-strained half Heuslers.
The notation introduced in Eq.~\eqref{eq_notation_tensors_4} is used.
Arrows indicate tensors that may be replaced by others due to
permutations
of indices, while tensors indicated by \eqref{eq_even_shear_lindep}
may be replaced by others due to Eq.~\eqref{eq_even_shear_lindep}.
}
\label{tab_tensors_even_shear}
\begin{ruledtabular}
\begin{tabular}{c|c|c||c|c|c|}
\#
&$\chi^{(\rm p\#)}$&Note
&\#
&$\chi^{(\rm p\#)}$&Note
\\
\hline
1 & $\langle333\rangle$ &
&35 &$\langle12113\rangle+\langle21223\rangle$ &$\rightarrow 21$\\
\hline
2 &$\langle231\rangle+\langle132\rangle$ &
&36 &$\langle11113\rangle+\langle22223\rangle$ &$\rightarrow 22$\\
\hline
3 & $\langle321\rangle+\langle312\rangle$ &
&37 &$\langle13313\rangle+\langle23323\rangle$ &$\rightarrow 27$\\
\hline
4 &$\langle311\rangle+\langle322\rangle$ &
&38 &$\langle23133\rangle+\langle13233\rangle$ &$\rightarrow 14$\\
\hline
5 &$\langle131\rangle+\langle232\rangle$ &\eqref{eq_even_shear_lindep}
&39 &$\langle13133\rangle+\langle23233\rangle$ &$\rightarrow 27$\\
\hline
6 & $\langle213\rangle+\langle123\rangle$ &
&40 &$\langle21333\rangle+\langle12333\rangle$ &\eqref{eq_even_shear_lindep}\\
\hline
7 &$\langle113\rangle+\langle223\rangle$ &
&41 &$\langle11333\rangle+\langle22333\rangle$ &\eqref{eq_even_shear_lindep}\\
\hline
8 & $\langle21321\rangle+\langle12312\rangle$ &
&42 &$\langle32221\rangle+\langle31112\rangle$& \\
\hline
9 &$\langle22321\rangle+\langle11312\rangle$ &
&43 &$\langle31221\rangle+\langle32112\rangle$&  \\
\hline
10 &$\langle21131\rangle+\langle12232\rangle$ &
&44 &$\langle32121\rangle+\langle31212\rangle$ &$\rightarrow 43$\\
\hline
11 & $\langle22131\rangle+\langle11232\rangle$ &
&45 &$\langle31121\rangle+\langle32212\rangle$&$\rightarrow 42$\\
\hline
12 & $\langle21231\rangle+\langle12132\rangle$ &$\rightarrow 8$
&46 &$\langle33321\rangle+\langle33312\rangle$ &\eqref{eq_even_shear_lindep}\\
\hline
13 &$\langle22231\rangle+\langle11132\rangle$ &$\rightarrow 9$
&47 &$\langle32211\rangle+\langle31122\rangle$ &$\rightarrow 43$\\
\hline
14 & $\langle23331\rangle+\langle13332\rangle$ &
&48 &$\langle31211\rangle+\langle32122\rangle$ &$\rightarrow 42$\\
\hline
15 &$\langle13221\rangle+\langle23112\rangle$ &
&49 &$\langle32111\rangle+\langle31222\rangle$ &\eqref{eq_even_shear_lindep}\\
\hline
16 &$\langle13121\rangle+\langle23212\rangle$ &
&50 &$\langle31111\rangle+\langle32222\rangle$ &\eqref{eq_even_shear_lindep}\\
\hline
17 & $\langle12321\rangle+\langle21312\rangle$ &$\rightarrow 8$
&51 &$\langle33311\rangle+\langle33322\rangle$ &\eqref{eq_even_shear_lindep}\\
\hline
18 &$\langle11321\rangle+\langle22312\rangle$ &$\rightarrow 9$
&52 &$\langle33231\rangle+\langle33132\rangle$ &$\rightarrow 46$\\
\hline
19 & $\langle13211\rangle+\langle23122\rangle$ &$\rightarrow 16$
&53 &$\langle33131\rangle+\langle33232\rangle$& $\rightarrow 51$\\
\hline
20 &$\langle13111\rangle+\langle23222\rangle$ &
&54 &$\langle32331\rangle+\langle31332\rangle$&\eqref{eq_even_shear_lindep}  \\
\hline
21 &$\langle12311\rangle+\langle21322\rangle$ &
&55 &$\langle31331\rangle+\langle32332\rangle$ &\eqref{eq_even_shear_lindep}\\
\hline
22 & $\langle11311\rangle+\langle22322\rangle$ &
&56 &$\langle33213\rangle+\langle33123\rangle$&$\rightarrow 46$\\
\hline
23 & $\langle12231\rangle+\langle21132\rangle$ &$\rightarrow 8$
&57 &$\langle33113\rangle+\langle33223\rangle$ &$\rightarrow 51$\\
\hline
24 &$\langle11231\rangle+\langle22132\rangle$ &$\rightarrow 9$
&58 &$\langle32313\rangle+\langle31323\rangle$ &$\rightarrow 54$\\
\hline
25 & $\langle12131\rangle+\langle21232\rangle$ &$\rightarrow 21$
&59 &$\langle31313\rangle+\langle32323\rangle$ &$\rightarrow 55$\\
\hline
26 &$\langle11131\rangle+\langle22232\rangle$ &$\rightarrow 22$
&60 &$\langle32133\rangle+\langle31233\rangle$ &$\rightarrow 54$\\
\hline
27 &$\langle13331\rangle+\langle23332\rangle$ &\eqref{eq_even_shear_lindep}
&61 &$\langle31133\rangle+\langle32233\rangle$ &$\rightarrow 55$\\
\hline
28 & $\langle21113\rangle+\langle12223\rangle$ &$\rightarrow 10$
&62 &$\langle33333\rangle$ &\eqref{eq_even_shear_lindep}\\
\hline
29 &$\langle22113\rangle+\langle11223\rangle$ &$\rightarrow 11$
&63 &$\langle23111\rangle+\langle13222\rangle$ &\eqref{eq_even_shear_lindep}\\
\hline
30 & $\langle21213\rangle+\langle12123\rangle$ &$\rightarrow 8$
&64 &$\langle23211\rangle+\langle13122\rangle$&$\rightarrow 15$ \\
\hline
31 &$\langle22213\rangle+\langle11123\rangle$ &$\rightarrow 9$
&65 &$\langle21311\rangle+\langle12322\rangle$& $\rightarrow 10$ \\
\hline
32 &$\langle23313\rangle+\langle13323\rangle$ &$\rightarrow 14$
&66 &$\langle22311\rangle+\langle11322\rangle$ &$\rightarrow 11$\\
\hline
33 & $\langle12213\rangle+\langle21123\rangle$ &$\rightarrow 8$
&67 &$\langle23121\rangle+\langle13212\rangle$&$\rightarrow 15$\\
\hline
34 & $\langle11213\rangle+\langle22123\rangle$ &$\rightarrow 9$
&68 &$\langle23221\rangle+\langle13112\rangle$&$\rightarrow 16$\\
\hline
\end{tabular}
\end{ruledtabular}
\end{threeparttable}

The corresponding torkance may be written as
%\bege\label{eq_even_shear_torkance_fit}
%t^{\rm even}_{ij}=\sum_{\#=1}^{7}\beta_{\#}\Xi_{in}\chi^{(\rm
%  p\#)}_{njk}\hat{M}_{k}+
%\sum_{\#=8}^{68}\beta_{\#}\Xi_{in}
%\chi^{({\rm p\#})}_{njklm}\hat{M}_{k}\hat{M}_{l}\hat{M}_{m}
%\ee
\bege\label{eq_even_shear_torkance_fit}
t^{\rm even}_{ij}=\Xi_{in}
\!\!
\left[
\sum_{\#=1}^{7}\beta_{\#}\chi^{(\rm
  p\#)}_{njk}+
\sum_{\#=8}^{68}\beta_{\#}
\chi^{({\rm p\#})}_{njklm}\hat{M}_{l}\hat{M}_{m}\!
\right]\!\!\hat{M}_{k},
\ee
where the matrix $\vn{\Xi}$ is defined in
Eq.~\eqref{eq_cross_op}. As discussed above, one may
set $\beta_{\#}=0$ for all tensors $\#$ indicated by an arrow 
in Table~\ref{tab_tensors_odd_shear}, i.e., $\beta_{12}=0, \beta_{13}=0,
\beta_{17}=0,\dots$. 

Due to the relations
\bege\label{eq_even_shear_lindep}
\begin{aligned}
&\Xi_{in}\chi^{(\rm
  p1)}_{njk}\hat{M}_{k}=
\Xi_{in}
[\chi^{({\rm p51})}_{njklm}
+
\chi^{({\rm p62})}_{njklm}
]
\hat{M}_{k}\hat{M}_{l}\hat{M}_{m}\\
&\Xi_{in}\chi^{(\rm
  p1)}_{njk}\hat{M}_{k}=-
\Xi_{in}\chi^{({\rm p5})}_{njklm}
\hat{M}_{k}\hat{M}_{l}\hat{M}_{m}\\
&0 =
\Xi_{in}
[\chi^{({\rm p27})}_{njklm}
+
\chi^{({\rm p62})}_{njklm}
]
\hat{M}_{k}\hat{M}_{l}\hat{M}_{m}\\
&0 =
\Xi_{in}
[\chi^{({\rm p16})}_{njklm}
+\frac{1}{2}
\chi^{({\rm p46})}_{njklm}
]
\hat{M}_{k}\hat{M}_{l}\hat{M}_{m}\\
&0=\Xi_{in}
\Bigl[
\chi^{(\rm
  p1)}_{njk}
+
[\chi^{({\rm p15})}_{njklm}
+
\chi^{({\rm p20})}_{njklm}
+
\chi^{({\rm p27})}_{njklm}
]
\hat{M}_{l}\hat{M}_{m}\Bigr]
\hat{M}_{k}\\
&0=\Xi_{in}
\Bigl[
\chi^{(\rm
  p2)}_{njk}
-
[\chi^{({\rm p14})}_{njklm}
+
\chi^{({\rm p63})}_{njklm}
-\frac{1}{2}
\chi^{({\rm p46})}_{njklm}
]
\hat{M}_{l}\hat{M}_{m}\Bigr]
\hat{M}_{k}\\
&0=\Xi_{in}
\Bigl[
\chi^{(\rm
  p3)}_{njk}
-
[\chi^{({\rm p42})}_{njklm}
+
\chi^{({\rm p49})}_{njklm}
+
\chi^{({\rm p54})}_{njklm}
]
\hat{M}_{l}\hat{M}_{m}\Bigr]
\hat{M}_{k}\\
&0=\Xi_{in}
\Bigl[
\chi^{(\rm
  p3)}_{njk}
+
[\chi^{({\rm p9})}_{njklm}
+
\chi^{({\rm p21})}_{njklm}
-
\chi^{({\rm p42})}_{njklm}\\
&-
\chi^{({\rm p49})}_{njklm}
]
\hat{M}_{l}\hat{M}_{m}\Bigr]
\hat{M}_{k}\\
&0=\Xi_{in}
\Bigl[
\chi^{(\rm
  p4)}_{njk}
-
[\chi^{({\rm p43})}_{njklm}
+
\chi^{({\rm p50})}_{njklm}
+
\chi^{({\rm p55})}_{njklm}
]
\hat{M}_{l}\hat{M}_{m}\Bigr]
\hat{M}_{k}\\
&0=\Xi_{in}
\Bigl[
\chi^{(\rm
  p4)}_{njk}
+
[\chi^{({\rm p8})}_{njklm}
+
\chi^{({\rm p22})}_{njklm}
-
\chi^{({\rm p43})}_{njklm}\\
&-
\chi^{({\rm p50})}_{njklm}
]
\hat{M}_{l}\hat{M}_{m}\Bigr]
\hat{M}_{k}\\
&0=\Xi_{in}
\Bigl[
\chi^{(\rm
  p6)}_{njk}
-
[\chi^{({\rm p10})}_{njklm}
+
\chi^{({\rm p21})}_{njklm}
+
\chi^{({\rm p40})}_{njklm}
]
\hat{M}_{l}\hat{M}_{m}\Bigr]
\hat{M}_{k}\\
&0=\Xi_{in}
\Bigl[
\chi^{(\rm
  p7)}_{njk}
-
[\chi^{({\rm p11})}_{njklm}
+
\chi^{({\rm p22})}_{njklm}
+
\chi^{({\rm p41})}_{njklm}
]
\hat{M}_{l}\hat{M}_{m}\Bigr]
\hat{M}_{k}\\
\end{aligned}
\ee
we may additionally set $\beta_{\#}=0$ 
in Eq.~\eqref{eq_even_shear_torkance_fit}
for $\#=5, 27, 40, 41, 46, 49, 50, 51, 54, 55, 62, 63$.
Thus, there are only 6 linearly independent polar tensors of rank 3
and 12 linearly independent polar tensors of rank 5 that need to be
considered in  Eq.~\eqref{eq_even_shear_torkance_fit}, i.e., 18 tensors in
total and 18 corresponding fitting parameters $\beta_{\#}$.

\bibliography{ptmnsb_dft_sot}

\end{document}